\pdfoutput=1

\documentclass[twocolumn,final]{svjour3}          % twocolumn

\journalname{arXiv preprint} % Insert the name of "your journal" with
\usepackage{graphicx}
\usepackage{mathptmx}      % use Times fonts if available on your TeX system
\usepackage[colorlinks=True,linkcolor=blue,citecolor=blue,anchorcolor=blue,urlcolor=blue]{hyperref}
\usepackage{amsmath} % for \pmb, \eqref and cases
\usepackage{amssymb}             % Extra maths symbols. Must be loaded before cleverref
\usepackage{natbib} % in bayesian analysis template ba-sample.tex
\usepackage{enumitem} % for custom enumerate
\usepackage[toc]{appendix}
\usepackage{mathtools} % for equalsdeff symbol \equiv
\usepackage[]{algorithm2e}
\usepackage[noabbrev]{cleveref}
\usepackage{booktabs} % for midrule, toprule etc
\usepackage{xcolor}  % needed for \textcolor in comment macros
\usepackage{bm}                  % For bold maths
\usepackage{layouts}  % for \\printinunitsof{in}\prntlen{\textwidth} - can remove in final version
\DeclareMathAlphabet{\mathcal}{OMS}{cmsy}{m}{n}
\newcommand{\dyPolyChord}{\href{https://github.com/ejhigson/dyPolyChord}{\texttt{dyPolyChord}}}
\newcommand{\dynesty}{\href{https://github.com/joshspeagle/dynesty}{\texttt{dynesty}}}
\newcommand{\perfectns}{\href{https://github.com/ejhigson/perfectns}{\texttt{perfectns}}}
\newcommand{\perfectnsurl}{\href{https://github.com/ejhigson/perfectns}{{https://github.com/ejhigson/perfectns}}}
\newcommand{\dyPolyChordurl}{\href{https://github.com/ejhigson/dyPolyChord}{{https://github.com/ejhigson/dyPolyChord}}}
\newcommand{\dynestyurl}{\href{https://github.com/joshspeagle/dynesty}{{https://github.com/joshspeagle/dynesty}}}
\newcommand{\PolyChord}{\texttt{PolyChord}}
\newcommand{\PolyChordtwo}{\texttt{PolyChord 2}}
\newcommand{\MultiNest}{\texttt{MultiNest}}
\newcommand{\nestcheck}{\texttt{nestcheck}}
\renewcommand{\d}[1]{\ensuremath{\operatorname{d}\!{#1}}}
\newcommand{\e}{\mathrm{e}}
\newcommand{\thcomp}[1]{\theta_{\hat{#1}}}
\newcommand{\ymean}[1]{\ensuremath{\overline{y (#1,\theta)}}}
\newcommand{\ndead}{n_\mathrm{dead}}
\newcommand{\btheta}{\ensuremath{\theta}}  % Do not make bold for this paper
\newcommand{\bmu}{\ensuremath{\mu}}        % Do not make bold for this paper
\newcommand{\thmean}[1]{\overline{\thcomp{#1}}}
\newcommand{\po}{\thmean{1}}
\newcommand{\std}{\ensuremath{\mathrm{St.Dev.}}}

\newcommand{\Z}{\mathcal{Z}}
\newcommand{\ninit}{n_\mathrm{init}}
\newcommand{\nbatch}{n_\mathrm{batch}}
\newcommand{\numrepeats}{\texttt{num\_repeats}}
\newcommand{\importance}{I}
\newcommand{\importancep}{\importance_\mathrm{param}}
\newcommand{\importancez}{\importance_\mathcal{Z}}
\begin{document}
\sloppy  % relax spacing between words to avoid them sticking into margin --- for more details see https://tex.stackexchange.com/questions/9107/how-can-i-make-my-text-never-go-over-the-right-margin-by-always-hyphenating-or-b

% Checking fonts:
% ---------------
% Use command here (will break compile) https://tex.stackexchange.com/questions/109703/how-to-determine-the-font-being-used-by-a-latex-document
% Output is form {encoding}{font}{series}{size} = e.g. \OT1/cmr/m/n/12 where m and n are just the normal series and shape
% For more details seek Section 4.3 of https://tug.org/texinfohtml/latex2e.html#Font-sizes
% cmr part is font (e.g. cmr = computer modern roman
% Font (=caption font): \OT1/ptm/m/n/10 (= times size 10)
% Caption font = General font
% Footnote font: \OT1/ptm/m/n/8.5
% Subplot caption font: \OT1/ptm/m/n/8.5
% Abstract font = general font
% Checking linewidth
% ------------------
% Just uncomment the below:
% textwidth: \printinunitsof{in}\prntlen{\textwidth}
% columnwidth: \printinunitsof{in}\prntlen{\columnwidth}
% linewidth: \printinunitsof{in}\prntlen{\linewidth}
% pdfpagewidth: \printinunitsof{in}\prntlen{\pdfpagewidth}
% textheight: \printinunitsof{in}\prntlen{\textheight}
% pdfpageheight: \printinunitsof{in}\prntlen{\pdfpageheight}
% lineheight: \printinunitsof{in}\prntlen{\textheight}

% Linewidth = textwidth = 6.39767in

\title{Dynamic nested sampling: an improved algorithm for parameter estimation and evidence calculation%
    \thanks{\textsuperscript{1} Dynamic nested sampling packages include:\newline
    \dyPolyChord{} (\dyPolyChordurl); Python, C\texttt{++} and Fortran likelihoods and priors, based on \PolyChord{}.\newline
    \dynesty{} (\dynestyurl); pure Python.\newline
    \perfectns{} (\perfectnsurl); pure Python, spherically symmetric likelihoods and priors only.}%
}
% \subtitle{Do you have a subtitle?\\ If so, write it here}

\titlerunning{Dynamic nested sampling}        % if too long for running head

\author{Edward Higson\textsuperscript{a,b} \and
Will Handley\textsuperscript{a,b} \and
Michael Hobson\textsuperscript{a} \and
Anthony Lasenby\textsuperscript{a,b}
}

\authorrunning{Higson et al.} % if too long for running head

\institute{\at{}\textsuperscript{a}Cavendish Astrophysics Group, University of Cambridge, UK \\
                \textsuperscript{b}Kavli Institute for Cosmology, University of Cambridge, UK \\
                \email{\href{mailto:e.higson@mrao.cam.ac.uk}{e.higson@mrao.cam.ac.uk}}           %  \\
}
\date{}  % The correct dates will be entered by the editor

\maketitle

% %%%%%%%%%%%%%%%%%%%%%%%%%%%%%%%%%%%%%%%%%%%%%%%%%%
% DNS paper body starts here
% %%%%%%%%%%%%%%%%%%%%%%%%%%%%%%%%%%%%%%%%%%%%%%%%%%

\begin{abstract}
We introduce dynamic nested sampling: a generalisation of the nested sampling algorithm in which the number of ``live points'' varies to allocate samples more efficiently.
In empirical tests the new method significantly improves calculation accuracy compared to standard nested sampling with the same number of samples; this increase in accuracy is equivalent to speeding up the computation by factors of up to $\sim72$ for parameter estimation and $\sim7$ for evidence calculations.
We also show that the accuracy of both parameter estimation and evidence calculations can be improved simultaneously.
In addition, unlike in standard nested sampling, more accurate results can be obtained by continuing the calculation for longer.
Popular standard nested sampling implementations can be easily adapted to perform dynamic nested sampling, and several dynamic nested sampling software packages are now publicly available.\!\footnotemark{}
\end{abstract}

\section{Introduction}

Nested sampling \citep{Skilling2006} is a numerical method for Bayesian computation which simultaneously provides both posterior samples and Bayesian evidence estimates.
The approach is closely related to Sequential Monte Carlo (SMC) \citep{Salomone2018} and rare event simulation \citep{Walter2017}.
The original development of the nested sampling algorithm was motivated by evidence calculation, but the \MultiNest{} \citep{Feroz2008,Feroz2009,Feroz2013} and \PolyChord{} \citep{Handley2015a,Handley2015b} software packages are now extensively used for parameter estimation from posterior samples \citep[such as in][]{DESCollaboration2017}.
Nested sampling performs well compared to Markov chain Monte Carlo (MCMC)-based parameter estimation for multi-modal and degenerate posteriors due to its lack of a thermal transition property and the relatively small amount of problem-specific tuning required; for example there is no need to specify a proposal function.
Furthermore, \PolyChord{} is well suited to high-dimensional parameter estimation problems due to its slice sampling-based implementation.

Nested sampling explores the posterior distribution by maintaining a set of samples from the prior, called {\em live points}, and iteratively updating them subject to the constraint that new samples have increasing likelihoods.
Conventionally a fixed number of live points is used; we term this {\em standard nested sampling}.
In this case the expected fractional shrinkage of the prior volume remaining is the same at each step, and as a result many samples are typically taken from regions of the prior that are remote from the bulk of the posterior.
The allocation of samples in standard nested sampling is set by the likelihood and the prior, and cannot be changed depending on whether calculating the evidence or obtaining posterior samples is the primary goal.

We propose modifying the nested sampling algorithm by dynamically varying the number of live points in order to maximise the accuracy of a calculation for some number of posterior samples, subject to practical constraints.
We term this more general approach {\em dynamic nested sampling}, with standard nested sampling representing the special case where the number of live points is constant.
Dynamic nested sampling is particularly effective for parameter estimation, as standard nested sampling typically spends most of its computational effort iterating towards the posterior peak.
This produces posterior samples with negligible weights which make little contribution to parameter estimation calculations, as discussed in our previous analysis of sampling errors in nested sampling parameter estimation \citep{Higson2017a}.
We also achieve significant improvements in the accuracy of evidence calculations, and show both evidence and parameter estimation can be improved simultaneously.
Our approach can be easily incorporated into existing standard nesting sampling software; we have created the \dyPolyChord{} package \citep{Higson2018dypolychord} for performing dynamic nested sampling using \PolyChord{}.

In this paper we demonstrate the advantages of dynamic nested sampling relative to the popular standard nested sampling algorithm in a range of empirical tests.
A detailed comparison of nested sampling with alternative methods such as MCMC-based parameter estimation and thermodynamic integration is beyond the current scope --- for this we refer the reader to \citet{Allison2014}, \citet{Murray2007} and \citet{Feroz2008thesis}.

The paper proceeds as follows: \Cref{sec:background} contains background on nested sampling, and \Cref{sec:vary_nlive} establishes useful results about the effects of varying the number of live points.
Our dynamic nested sampling algorithm for increasing efficiency in general nested sampling calculations is presented in \Cref{sec:dns}; its accurate allocation of live points for {\em a priori\/} unknown posterior distributions is illustrated in \Cref{fig:nlive_gaussian}.
We first test dynamic nested sampling in the manner described by \citet{Keeton2011}, using analytical cases where one can obtain uncorrelated samples from the prior space within some likelihood contour using standard techniques.
We term the resulting procedure {\em perfect nested sampling\/} (in both standard and dynamic versions), and use it to compare the performance of dynamic and standard nested sampling in a variety of cases without software-specific effects from correlated samples or prohibitive computational costs.
These tests were performed with our \perfectns{} package \citep{Higson2018perfectns} and are described in \Cref{sec:numerical_tests}, which includes a discussion of the effects of likelihood, priors and dimensionality on the improvements from dynamic nested sampling.
In particular we find large efficiency gains for high-dimensional parameter estimation problems.

\Cref{sec:practical_problems} discusses applying dynamic nested sampling to challenging posteriors, in which results from nested sampling software may include implementation-specific effects from correlations between samples \citep[see][for a detailed discussion]{Higson2018a}.
We describe the strengths and weaknesses of dynamic nested sampling compared to standard nested sampling in such cases.
This section includes numerical tests with a multimodal Gaussian mixture model and a practical signal reconstruction problem using \dyPolyChord{}.
We find that dynamic nested sampling also produces significant accuracy gains for these more challenging posteriors, and that it is able to reduce implementation-specific effects compared to standard nested sampling.

\subsection{Other related work}

Other variants of nested sampling include diffusive nested sampling \citep{Brewer2011} and superposition enhanced nested sampling \citep{Martiniani2014}, which have been implemented as stand alone software packages.
In particular, dynamic nested sampling shares some similarities with \texttt{DNest4} \citep{Brewer2016}, in which diffusive nested sampling is followed by additional sampling targeting regions of high posterior mass.
However dynamic nested sampling differs from these alternatives as, like standard nested sampling, it only requires drawing samples within hard likelihood constraints.
As a result dynamic nested sampling can be used to improve the efficiency of popular standard nested sampling implementations such as \MultiNest{} (rejection sampling), \PolyChord{} (slice sampling) and constrained Hamiltonian nested sampling \citep{Betancourt2011} while maintaining their strengths in sampling degenerate and multimodal distributions.

It has been shown that efficiency can be greatly increased using nested importance sampling \citep{Chopin2010} or by performing nested sampling using an auxiliary prior which approximates the posterior as described in \citet{Cameron2014}. However, the efficacy of these approaches is contingent on having adequate knowledge of the posterior (either before the algorithm is run, or by using the results of previous runs). As such, the speed increase on {\em a priori\/} unknown problems is generally lower than might be suggested by toy examples.

Dynamic nested sampling is similar in spirit to the adaptive schemes for thermodynamic integration introduced by \citet{Hug2016} and \citet{Friel2014}, as each involves an initial run followed by additional targeted sampling using an estimated error criteria.
Furthermore, dynamically weighting sampling in order to target regions of higher posterior mass has also been used in the statistical physics literature, such as in multi-canonical sampling \citep[see for example][]{Okamoto2004}.

\section{Background: the nested sampling algorithm}\label{sec:background}

We now give a brief description of the nested sampling algorithm following \citet{Higson2017a} and set out our notation; for more details see \citet{Higson2017a} and \citet{Skilling2006}.
For theoretical treatments of nested sampling's convergence properties, see \citet{Keeton2011,Skilling2009,Walter2017,Evans2007}.

For a given likelihood $\mathcal{L}(\btheta)$ and prior $\pi(\btheta)$, nested sampling is a method for simultaneously computing the Bayesian evidence
\begin{equation}
    \mathcal{Z}
    =
    \int \mathcal{L}(\btheta) \pi (\btheta)\d{\btheta}
    \label{equ:Z_definition}
\end{equation}
and samples from the posterior distribution
\begin{equation}
    \mathcal{P}(\btheta) = \frac{\mathcal{L}(\btheta) \pi(\btheta)}{\mathcal{Z}}.
    \label{equ:parameter_estimation}
\end{equation}
The algorithm begins by sampling some number of {\em live points\/} randomly from the prior $\pi(\btheta)$.
In standard nested sampling, at each iteration $i$ the point with the lowest likelihood $\mathcal{L}_i$ is replaced by a new point sampled from the region of prior with likelihood $\mathcal{L}(\btheta)>\mathcal{L}_i$ and the number of live points remains constant throughout.
This process is continued until some termination condition is met, producing a list of samples (referred to as {\em dead points\/}) which --- along with any remaining live points --- can then be used for evidence and parameter estimation.
We term the finished nested sampling process a {\em run}.

\begin{figure}
    \centering
    \includegraphics[width=\linewidth]{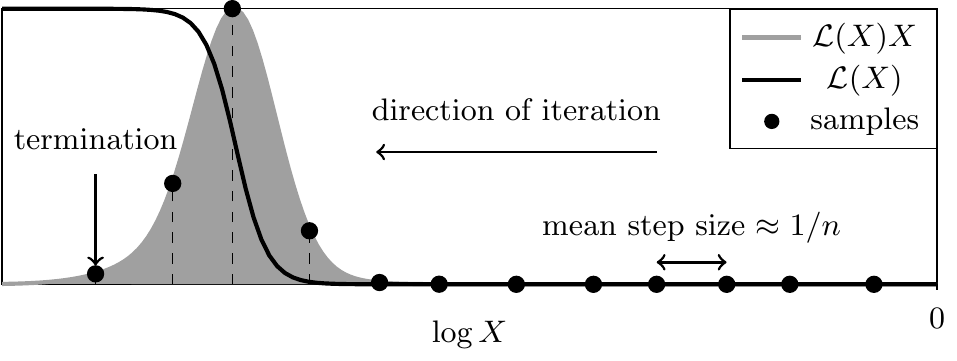}
    \caption{A schematic illustration of standard nested sampling with a constant number of live points $n$ reproduced from \citet{Higson2017a}.
$\mathcal{L}(X)X$ shows the relative posterior mass, the bulk of which is contained in some small fraction of the prior.
Most of the samples in the diagram are in $\log X$ regions with negligible posterior mass, as is typically the case in standard nested sampling.%
}\label{fig:ns_evidence}
\end{figure}

Nested sampling calculates the evidence~\eqref{equ:Z_definition} as a one-dimensional integral
\begin{equation}
    \mathcal{Z}=\int_0^1 \mathcal{L}(X) \d{X},
    \label{equ:Z(X)}
\end{equation}
where $X(\mathcal{L})$ is the fraction of the prior with likelihood greater than $\mathcal{L}$ and $\mathcal{L}(X)\equiv X^{-1}(\mathcal{L})$.
The prior volumes $X_i$ corresponding to the dead points $i$ are unknown but can be modelled statistically as $X_i = t_i X_{i-1}$, where $X_0 = 1$.
For a given number of live points $n$, each shrinkage ratio $t_i$ is independently distributed as the largest of $n$ random variables from the interval $[0,1]$ and so \citep{Skilling2006}:
\begin{equation}
    P(t_i)            = n t_i^{n-1}, \qquad
    \mathrm{E}[\log t_i ]        = -\frac{1}{n},   \qquad
    \mathrm{Var}[\log t_i ]    = \frac{1}{n^2}.
    \label{equ:dist_t}
\end{equation}
In standard nested sampling the number of live points $n$ is some constant value for all $t_i$ --- the iteration of the algorithm in this case is illustrated schematically in \Cref{fig:ns_evidence}.

\subsubsection*{Evidence estimation}\label{sec:sampling_evidence_error}
Nested sampling calculates the evidence~\eqref{equ:Z(X)} as a quadrature sum over the dead points
\begin{equation}
    \mathcal{Z}(\mathbf{t}) \approx \sum_{i \in \mathrm{dead}} \mathcal{L}_i w_i(\mathbf{t}),
    \label{equ:ztot}
\end{equation}
where $\mathbf{t}=\{t_1,t_2,\dots,t_{\ndead}\}$ are the unknown set of shrinkage ratios for each dead point and each $t_i$ is an independent random variable with distribution~\eqref{equ:dist_t}.
If required any live points remaining at termination can also be included.
The $w_i$ are appropriately chosen quadrature weights; we use the trapezium rule such that $w_i(\mathbf{t})=\frac{1}{2}(X_{i-1}(\mathbf{t})-X_{i+1}(\mathbf{t}))$, where $X_i(\mathbf{t}) = \prod^i_{k=0} t_k$.
Given that the shrinkage ratios $\mathbf{t}$ are {\em a priori\/} unknown, one typically calculates an expected value and error on the evidence~\eqref{equ:ztot} using~\eqref{equ:dist_t}.
The dominant source of error in evidence estimates from perfect nested sampling is the statistical variation in the unknown volumes of the prior ``shells'' $w_i(\mathbf{t})$.

\subsubsection*{Parameter estimation}\label{sec:sampling_parameter_error} 

Nested sampling parameter estimation uses the dead points, and if required the remaining live points at termination, to construct a set of posterior samples with weights proportional to their share of the posterior mass:
\begin{equation}
    p_i(\mathbf{t})=\frac{w_i(\mathbf{t})\mathcal{L}_i}{\sum_i w_i(\mathbf{t})\mathcal{L}_i}=\frac{w_i(\mathbf{t})\mathcal{L}_i}{\mathcal{Z}(\mathbf{t})}.
    \label{equ:posterior_weight}
\end{equation}
Neglecting any implementation-specific effects, which are not present in perfect nested sampling, the dominant sampling errors in estimating some parameter or function of parameters $f(\btheta)$ come from two sources \citep{Higson2017a}:
\begin{enumerate}[label= (\roman*)]
    \item approximating the relative point weights $p_i(\mathbf{t})$ with their expectation $\mathrm{E}[p_i(\mathbf{t})]$ using~\eqref{equ:dist_t};\label{enu:w_error}
    \item approximating the mean value of a function of parameters over an entire iso-likelihood contour with its value at a single point $f(\btheta_i)$.\label{enu:sample_error}
\end{enumerate}

\subsubsection*{Combining and dividing nested sampling runs}\label{sec:divide}

\citet{Skilling2006} describes how several standard nested sampling runs $r=1,2,\dots$ with constant live points $n^{(r)}$ may be combined simply by merging the dead points and sorting by likelihood value.
The combined sequence of dead points is equivalent to a single nested sampling run with $n_\mathrm{combined}=\sum_r n^{(r)}$ live points.

\citet{Higson2017a} gives an algorithm for the reverse procedure: decomposing a nested sampling run with $n$ live points into a set of $n$ valid nested sampling runs, each with 1 live point.
These single live point runs, which we term {\em threads}, are the smallest unit from which valid nested sampling runs can be constructed and will prove useful in developing dynamic nested sampling.

\section{Variable numbers of live points}\label{sec:vary_nlive}

Before presenting our dynamic nested sampling algorithm in \Cref{sec:dns}, we first establish some basic results for a nested sampling run in which the number of live points varies.
Such runs are valid as successive shrinkage ratios $t_i$ are independently distributed \citep{Skilling2006}.
For now we assume the manner in which the number of live points changes is specified in advance; adaptive allocation of samples is considered in the next section.

Let us define $n_i$ as the number of live points present for the prior shrinkage ratio $t_i$ between dead points $i-1$ and $i$.\footnote{In order for~\eqref{equ:dist_t} to be valid, the number of live points must remain constant across the shrinkage ratios $t_i$ between successive dead points. We therefore only allow the number of live points to change on iso-likelihood contours $\mathcal{L}(\btheta) = \mathcal{L}_i$ where a dead point $i$ is present.
This restriction has negligible effects for typical calculations, and is automatically satisfied by most nested sampling implementations.}
In this notation all information about the number of live points for a nested sampling run can be expressed as a list of numbers $\mathbf{n} = \{n_1, n_2, \dots, n_{\ndead}\}$ which correspond to the shrinkage ratios $\mathbf{t} = \{t_1,t_2,\dots,t_{\ndead}\}$.
Nested sampling calculations for variable numbers of live points differ from the constant live point case only in the use of different $n_i$ in calculating the distribution of each $t_i$ from~\eqref{equ:dist_t}.

\begin{figure}
	\centering
    \includegraphics[width=\linewidth]{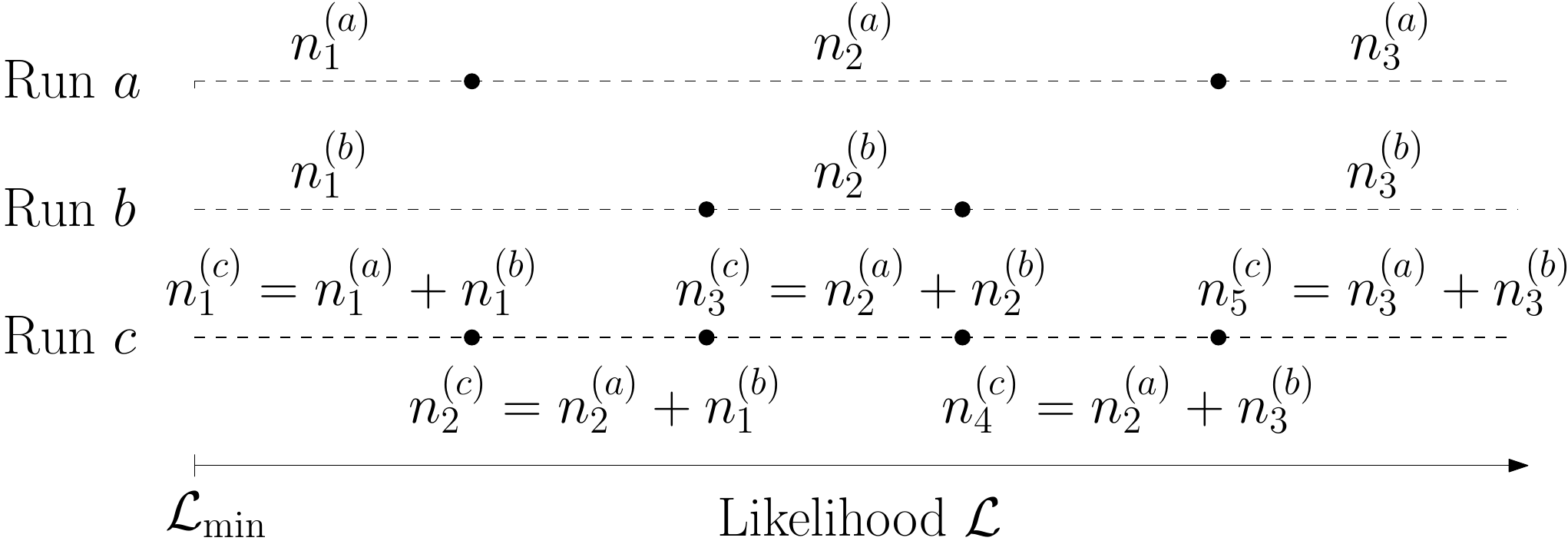}
    \caption{Combining nested sampling runs $a$ and $b$ with variable numbers of live points $\mathbf{n}^{(a)}$ and $\mathbf{n}^{(b)}$ into a single nested sampling run $c$; black dots show dead points arranged in order of increasing likelihood.
The number of live points in run $c$ at some likelihood equals the sum of the live points of run $a$ and run $b$ at that likelihood.
}\label{fig:combining_dynamic}
\end{figure}

\citet{Skilling2006}'s method for combining constant live point runs, mentioned in \Cref{sec:divide}, can be extended to accommodate variable numbers of live points by requiring that at any likelihood the live points of the combined run equals the sum of the live points of the constituent runs at that likelihood (this is illustrated in \Cref{fig:combining_dynamic}).
Variable live point runs can also be divided into their constituent threads using the algorithm in \citet{Higson2017a}. However, unlike for constant live point runs, the threads produced may start and finish part way through the run and there is no longer a single unique division into threads on iso-likelihood contours where the number of live points increases.
The technique for estimating sampling errors by resampling threads introduced in \citet{Higson2017a} can also be applied for nested sampling runs with variable numbers of live points (see Appendix~\ref{app:bootstrap} for more details), as can the diagnostic tests for correlated samples and missed modes described in \citet{Higson2018a}.

In addition, the variable live point framework provides a natural way to include the final set of live points remaining when a standard nested sampling run terminates in a calculation.
These are uniformly distributed in the region of the prior with $\mathcal{L}(\btheta) > \mathcal{L}_\mathrm{terminate}$, and can be treated as samples from a dynamic nested sampling run with the number of live points reducing by 1 as each of the points remaining after termination is passed until the final point $i$ has $n_i = 1$.
This allows the final live points of standard nested sampling runs to be combined with variable live point runs.

The remainder of this section analyses the effects of local variations in the number of live points on the accuracy of nested sampling evidence calculation and parameter estimation.
The dynamic nested sampling algorithm in \Cref{sec:dns} uses these results to allocate additional live points.

\subsection{Effects on calculation accuracy}\label{sec:optimum_w}

Nested sampling calculates the evidence $\mathcal{Z}$ as the sum of sample weights~\eqref{equ:ztot}; the dominant sampling errors are from statistically estimating shrinkage ratios $t_i$ which affect the weights of all subsequent points.
In Appendix~\ref{app:optimum_z_derivation} we show analytically that the reduction in evidence errors achieved by taking additional samples to increase the local number of live points $n_i$ is inversely proportional to $n_i$, and is approximately proportional to the evidence contained in point $i$ and all subsequent points.
This makes sense as the dominant evidence errors are from statistically estimating shrinkages $t_i$ which affect all points $j \ge i$.

In nested sampling parameter estimation, sampling errors come both from taking a finite number of samples in any region of the prior and from the stochastic estimation of their normalised weights $p_i$ from~\eqref{equ:posterior_weight}.
Typically standard nested sampling takes many samples with negligible posterior mass as illustrated in \Cref{fig:ns_evidence}; these make little contribution to estimates of parameters or to the accuracy of samples' normalised weights.
From~\eqref{equ:dist_t} the expected separation between points in $\log X$ (approximately proportional to the posterior mass they each represent) is $1/n_i$.
As a result, increasing the number of live points wherever the dead points' posterior weights $p_i \propto \mathcal{L}_i w_i$ are greatest distributes posterior mass more evenly among the samples.
This improves the accuracy of the statistically estimated weights $p_i$, and can dramatically increase the information content (Shannon entropy of the samples)
\begin{equation}
	H = \exp\left( - \sum_i p_i \log p_i \right),
    \label{equ:entropy}
\end{equation}
which is maximised for a given number of samples when the sample weights are equal.
Empirical tests of dynamic nested sampling show that increasing the number of live points wherever points have the highest $p_i \propto \mathcal{L}_i w_i$ works well as regards increasing parameter estimation accuracy for most calculations.

As the contribution of each sample $i$ to a parameter estimation problem for some quantity $f(\btheta)$ is dependent on $f(\btheta_i)$, the precise optimum allocation of live points is different for different quantities.
In most cases the relative weight $p_i$ of samples is a good approximation for their influence on a calculation, but for some problems much of the error may come from sampling $\log X$ regions containing a small fraction of the posterior mass but with extreme parameter values \citep[see Section 3.1 of][for diagrams illustrating this]{Higson2017a}.
Appendix~\ref{app:tuning} discusses estimating the importance of points to a specific parameter estimation calculation and using dynamic nested sampling to allocate live points accordingly.

\section{The dynamic nested sampling algorithm}\label{sec:dns}

This section presents our algorithm for performing nested sampling calculations with a dynamically varying number of live points to optimise the allocation of samples.

Since the distribution of posterior mass as a function of the likelihood is {\em a priori\/} unknown, we first approximate it by performing a standard nested sampling run with some small constant number of live points $\ninit$.
The algorithm then proceeds by iteratively calculating the range of likelihoods where increasing the number of live points will have the greatest effect on calculation accuracy, and generating an additional thread running over these likelihoods.
If required some $\nbatch$ additional threads can be generated at each step to reduce the number of times the importance must be calculated and the sampler restarted.
We find in empirical tests that using $\nbatch > 1$ has little effect on efficiency gains from dynamic nested sampling when the number of samples taken in each batch is small compared to the total number of samples in the run.

From the discussion in \Cref{sec:optimum_w} we define functions to measure the relative importance of a sample $i$ for evidence calculation
and parameter estimation respectively as
\begin{align}
    \importancez(i)
    &\propto                  
    \frac{\mathrm{E}[\mathcal{Z}_{\ge i}]}{n_i}, \quad \text{where} \, \mathcal{Z}_{\ge i} \equiv \sum_{k \ge i} \mathcal{L}_k w_k(\mathbf{t}),
    \label{equ:z_importance}
    \\
    \importancep(i)
    &\propto
    \mathcal{L}_i \,\, \mathrm{E}[w_i(\mathbf{t})].\label{equ:p_importance}
\end{align}
Alternatively~\eqref{equ:z_importance} can be replaced with the more complex expression~\eqref{equ:exact_z_importance} derived in Appendix~\ref{app:optimum_z_derivation}, although we find this typically makes little difference to results.
Modifying~\eqref{equ:p_importance} to optimise for estimation of a specific parameter or function of parameters is discussed in Appendix~\ref{app:tuning}.

The user specifies how to divide computational resources between evidence calculation and parameter estimation through an input goal $G \in [0,1]$, where $G=0$ corresponds to optimising for evidence calculation and $G=1$ optimises for parameter estimation.
The dynamic nested sampling algorithm calculates importance as a weighted sum of the points' normalised evidence and parameter estimation importances
\begin{equation}
    \importance(G, i)
    =
    (1-G) \frac{\importancez(i)}{\sum_j \importancez(j)}
    +
    G \frac{\importancep(i)}{\sum_j \importancep(j)}.
    \label{equ:importance}
\end{equation}

The likelihood range in which to run an additional thread is chosen by finding all points with importance greater than some fraction $f$ of the largest importance.
Choosing a smaller fraction makes the threads added longer and reduces the number of times the importance must be recalculated, but can also cause the number of live points to plateau for regions with importance greater than that fraction of the maximum importance (see the discussion of~\Cref{fig:nlive_gaussian} in the next section for more details).
We use $f = 0.9$ for results in this paper, but find empirically that using slightly higher or lower values make little difference to results.
To ensure any steep or discontinuous increases in the likelihood $\mathcal{L}(X)$ are captured we find the first point $j$ and last point $k$ which meet this condition, then generate an additional thread starting at $\mathcal{L}_{j-1}$ and ending when a point is sampled with likelihood greater than $\mathcal{L}_{k+1}$.
If $j$ is the first dead point, threads which initially sample the whole prior are generated.
If $k$ is the final dead point then the thread will stop when a sample with likelihood greater than $\mathcal{L}_k$ is found.\footnote{We find empirically that one additional point per thread is sufficient to reach higher likelihoods if required. This is because typically there are many threads, and for each thread (which has only one live point) the expected shrinkage between samples~\eqref{equ:dist_t} of $E[\log t_i] = -1$ is quite large.}
This allows the new thread to continue beyond $\mathcal{L}_k$, meaning dynamic nested sampling iteratively explores higher likelihoods when this is the most effective use of samples.

Unlike in standard nested sampling, more accurate dynamic nested sampling results can be obtained simply by continuing the calculation for longer.
The user must specify a condition at which to stop dynamically adding threads, such as when fixed number of samples has been taken or some desired level of accuracy has been achieved.
Sampling errors on evidence and parameter estimation calculations can be estimated from the dead points at any stage using the method described in \citet{Higson2017a}.
We term these {\em dynamic termination conditions\/} to distinguish them from the type of termination conditions used in standard nested sampling.
Our dynamic nested sampling algorithm is presented more formally in Algorithm~\ref{alg:dns}.

\begin{algorithm}\SetAlgoLined{}
\SetKwData{Left}{left}\SetKwData{This}{this}\SetKwData{Up}{up}
\SetKwFunction{Union}{Union}\SetKwFunction{FindCompress}{FindCompress}
\SetKwInOut{Input}{Input}\SetKwInOut{Output}{Output}\SetKwInOut{op}{Other parameters}
    \Output{Samples and live points information $\mathbf{n}$.}
    \Input{Goal $G$, $\ninit$, dynamic termination condition.}
    \BlankLine{}
    \PrintSemicolon{}
    Generate a nested sampling run with a constant number of live points $\ninit$\;
    \While{dynamic termination condition not satisfied}{%
        recalculate importance $\importance(G, i)$ of all points\;
        find first point $j$ and last point $k$ with importance of greater than some fraction $f$ (we use $f=0.9$) of the largest importance\;
        generate an additional thread (or alternatively $n_\mathrm{batch}$ additional threads) starting at $\mathcal{L}_{j-1}$ and ending with the first sample taken with likelihood greater than $\mathcal{L}_{k+1}$\footnotemark{}\;
    }
    \caption{Dynamic nested sampling.}\label{alg:dns}
\end{algorithm}%
\footnotetext{If $k$ is the final dead point, the additional thread terminates after the first point with likelihood greater than $\mathcal{L}_k$.}

\subsection{Software implementation}

Since dynamic nested sampling only requires the ability to sample from the prior within a hard likelihood constraint, implementations and software packages developed for standard nested sampling can be easily adapted to perform dynamic nested sampling.
We demonstrate this with the \dyPolyChord{} package, which performs dynamic nested sampling using \PolyChord{} and is compatible with Python, C\texttt{++} and Fortran likelihoods.

\PolyChord{} was designed before the creation of the dynamic nested sampling algorithm, and is not optimized to quickly resume the nested sampling process at an arbitrary point to add more threads.
\dyPolyChord, which performs nested sampling with \PolyChord{}, minimises the computational overhead from saving and resuming by using Algorithm~\ref{alg:dypolychord} --- a modified version of Algorithm~\ref{alg:dns} described in Appendix~\ref{app:dyPolyChord}.
After the initial exploratory run with $\ninit$ live points, Algorithm~\ref{alg:dypolychord} calculates a dynamic allocation of live points and then generates more samples in a single run without recalculating point importances.
This means only the initial run provides information on where to place samples, and as a result the allocation of live points is slightly less accurate and a higher value of $\ninit$ is typically needed.

Dynamic nested sampling will be incorporated in the forthcoming \PolyChordtwo{} software package, which is currently in development and is designed for problems of up to $\sim 1,000$ dimensions --- dynamic nested sampling can provide very large improvements in the accuracy of such high-dimensional problems, as shown by the numerical tests in the next section.
Furthermore, we anticipate reloading a past iteration $i$ of a \PolyChordtwo{} nested sampling run in order to add additional threads will be less computationally expensive than a single likelihood call for many problems.
Nevertheless, it is often more efficient for dynamic nested sampling software to generate additional threads in selected likelihood regions in batches rather than one at a time; this approach is used in the \dynesty{}\footnote{See \dynestyurl{} for more information.} dynamic nested sampling package.

\section{Numerical tests with perfect nested sampling}\label{sec:numerical_tests}

In the manner described by \citet{Keeton2011} we first consider spherically symmetric test cases; here one can perform {\em perfect nested sampling}, as perfectly uncorrelated samples from the prior space within some iso-likelihood contour can be found using standard techniques.
Results from nested sampling software used for practical problems may include additional uncertainties from imperfect sampling within a likelihood contour that are specific to a given implementation --- we discuss these in \Cref{sec:practical_problems}.
The tests in this section were run using our \perfectns{} package.

Perfect nested sampling calculations depend on the likelihood $\mathcal{L}(\btheta)$ and prior $\pi(\btheta)$ only through the distribution of posterior mass $\mathcal{L}(X)$ and the distribution of parameters on iso-likelihood contours $P(f(\btheta)|\mathcal{L}(\btheta)=\mathcal{L}(X))$, each of which are functions of both $\mathcal{L}(\btheta)$ and $\pi(\btheta)$ \citep{Higson2017a}.
We therefore empirically test dynamic nested sampling using likelihoods and priors with a wide range of distributions of posterior mass, and consider a variety of functions of parameters $f(\btheta)$ in each case.

\begin{figure*}
	\centering
    \includegraphics{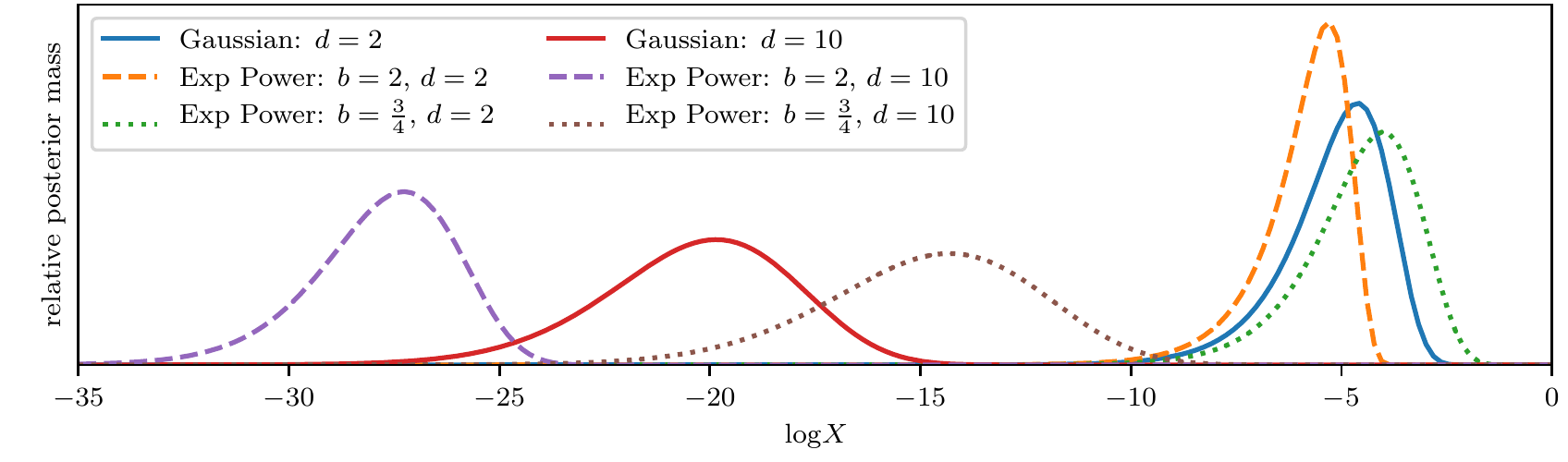}
    \caption{Relative posterior mass ($\propto \mathcal{L}(X)X$) as a function of $\log X$ for Gaussian likelihoods~\eqref{equ:gaussian} and exponential power likelihoods~\eqref{equ:exp_power} with $b=2$ and $b=\frac{3}{4}$. Each has a Gaussian prior~\eqref{equ:gaussian_prior} with $\sigma_\pi=10$.
The lines are scaled so that the area under each of them is equal.}\label{fig:an_w}
\end{figure*}

We first examine perfect nested sampling of $d$-dimensional spherical unit Gaussian likelihoods centred on the origin
\begin{equation}\label{equ:gaussian}
    \mathcal{L}(\btheta) = {(2 \pi)}^{-d/2} \e^{-{|\btheta|}^2 / 2}.
\end{equation}
For additional tests using distributions with lighter and heavier tails we use $d$-dimensional exponential power likelihoods
\begin{equation}\label{equ:exp_power}
    \mathcal{L}(\btheta) = \frac{d\, \Gamma(\frac{d}{2})}{{\pi}^{\frac{d}{2}} 2^{1+\frac{1}{2b}} \Gamma(1+\frac{n}{2b})} \mathrm{e}^{-{|\btheta|}^{2b} / 2},
\end{equation}
where $b=1$ corresponds to a $d$-dimensional Gaussian~\eqref{equ:gaussian}.
All tests use $d$-dimensional co-centred spherical Gaussian priors
\begin{equation}\label{equ:gaussian_prior}
    \pi(\btheta) = {(2 \pi \sigma_\pi^2)}^{-d/2} \e^{-{|\btheta|}^2 / 2 \sigma_\pi^2}.
\end{equation}
The different distributions of posterior mass in $\log X$ for~\eqref{equ:gaussian} and~\eqref{equ:exp_power} with dimensions $d$ are illustrated in \Cref{fig:an_w}.

In tests of parameter estimation we denote the first component of the $\btheta$ vector as $\thcomp{1}$, although by symmetry the results will be the same for any component.
$\po$ is the mean of the posterior distribution of $\thcomp{1}$, and the one-tailed $Y\%$ upper credible interval $\mathrm{C.I.}_{Y\%}(\thcomp{1})$ is the value $\thcomp{1}^\ast$ for which $P(\thcomp{1}<\thcomp{1}^\ast|\mathcal{L},\pi)=Y/100$.

Tests of dynamic nested sampling terminate after a fixed number of samples, which is set such that they use similar or slightly smaller numbers of samples than the standard nested sampling runs we compare them to.
Dynamic runs have $\ninit$ set to 10\% of the number of live points used for the standard runs.
Standard nested sampling runs use the termination conditions described by \citet[][Section 3.4]{Handley2015b}, stopping when the estimated evidence contained in the live points is less than $10^{-3}$ times the evidence contained in dead points (the default value used in \PolyChord).
This is an appropriate termination condition for nested sampling parameter estimation \citep{Higson2017a}, but if only the evidence is of interest then stopping with a larger fraction of the posterior mass remaining will have little effect on calculation accuracy.

The increase in computational efficiency from our method can be calculated by observing that nested sampling calculation errors are typically proportional to the square root of the computational effort applied \citep{Skilling2006,Higson2017a}, and that the number of samples produced is approximately proportional to the computational effort.
The increase in efficiency (computational speedup) from dynamic nested sampling over standard nested sampling for runs containing approximately the same number of samples on average can therefore be estimated from the variation of results as
\begin{equation}
    \mathrm{efficiency\,gain} = \frac{\mathrm{Var}\left[\mathrm{standard\,NS\,results}\right]}{\mathrm{Var}\left[\mathrm{dynamic\,NS\,results}\right]}.\label{equ:efficiency_gain}
\end{equation}
Here the numerator is the variance of the calculated values of some quantity (such as the evidence or the mean of a parameter) from a number of standard nested nested sampling runs, and the denominator is the variance of the calculated values of the same quantity from a number of dynamic nested sampling runs.
When the two methods use different numbers of samples on average,~\eqref{equ:efficiency_gain} can be replaced with
\begin{equation}
    \mathrm{efficiency\,gain} = \frac{\mathrm{Var}\left[\mathrm{standard\,NS\,results}\right]}{\mathrm{Var}\left[\mathrm{dynamic\,NS\,results}\right]}
    \times
    \frac{\overline{N_\mathrm{samp,sta}}}{\overline{N_\mathrm{samp,dyn}}},% \frac{\overbar{N_\mathrm{samp,std}}}{\overbar{N_\mathrm{samp,std}}}
    \label{equ:efficiency_gain_nsamp}
\end{equation}
where the additional term is the ratio of the mean number of samples produced by the standard and dynamic nested sampling runs.

\subsection{10-dimensional Gaussian example}

\begin{figure*}
	\centering
    \includegraphics{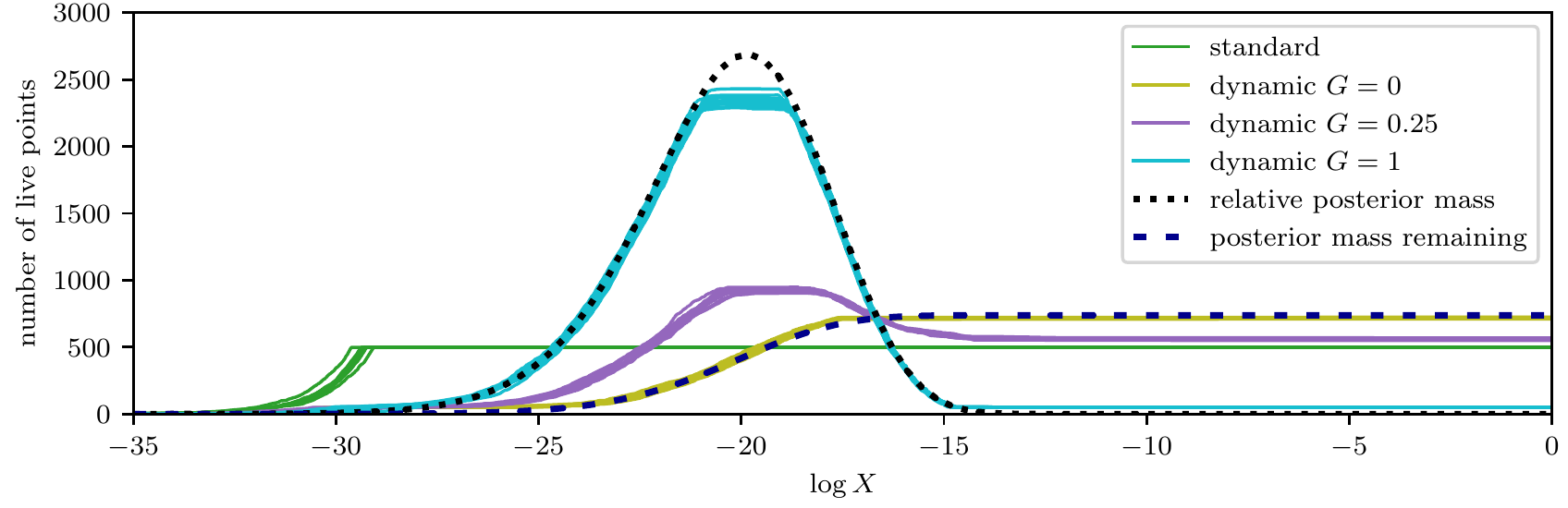}
    \caption{Live point allocation for a 10-dimensional Gaussian likelihood~\eqref{equ:gaussian} with a Gaussian prior~\eqref{equ:gaussian_prior} and $\sigma_\pi = 10$.
Solid lines show the number of live points as a function of $\log X$ for 10 standard nested sampling runs with $n=500$, and 10 dynamic nested sampling runs with $\ninit=50$, a similar number of samples and different values of $G$.
The dotted and dashed lines show the relative posterior mass $\propto \mathcal{L}(X)X$ and the posterior mass remaining $\propto \int_{-\infty}^X \mathcal{L}(X')X' \d{X'}$ at each point in $\log X$; for comparison these lines are scaled to have the same area under them as the average of the number of live point lines.
Standard nested sampling runs include the final set of live points at termination, which are modeled using a decreasing number of live points as discussed in \Cref{sec:vary_nlive}.
Similar diagrams for exponential power likelihoods~\eqref{equ:exp_power} with $b=2$ and $b=\frac{3}{4}$ are presented in \Cref{fig:nlive_exp_power_2,fig:nlive_exp_power_0_75} in Appendix~\ref{app:exp_power_add_tests}.
}\label{fig:nlive_gaussian}
\end{figure*}

We begin by testing dynamic nested sampling on a 10-dimensional Gaussian likelihood~\eqref{equ:gaussian} with a Gaussian prior~\eqref{equ:gaussian_prior} and $\sigma_\pi = 10$.
\Cref{fig:nlive_gaussian} shows the relative allocation of live points as a function of $\log X$ for standard and dynamic nested sampling runs.
The dynamic nested sampling algorithm (Algorithm~\ref{alg:dns}) can accurately and consistently allocate live points, as can be seen by comparison with the analytically calculated distribution of posterior mass and posterior mass remaining.
Dynamic nested sampling live point allocations do not precisely match the distribution of posterior mass and posterior mass remaining in the $G=1$ and $G=0$ cases because they include the initial exploratory run with a constant $\ninit$ live points.
Furthermore as additional live points are added where the importance is more than $90\%$ of the maximum importance, the number of live points allocated by dynamic nested sampling is approximately constant for regions with importance of greater than $\sim 90\%$ of the maximum --- this can be clearly seen in \Cref{fig:nlive_gaussian} near the peak number of live points in the $G=1$ case.
Similar diagrams for exponential power likelihoods~\eqref{equ:exp_power} with $b=2$ and $b=\frac{3}{4}$ are provided in Appendix~\ref{app:exp_power_add_tests} (\Cref{fig:nlive_exp_power_2,fig:nlive_exp_power_0_75}), and show the allocation of live points is also accurate in these cases.

\begin{figure*}
	\centering
    \includegraphics{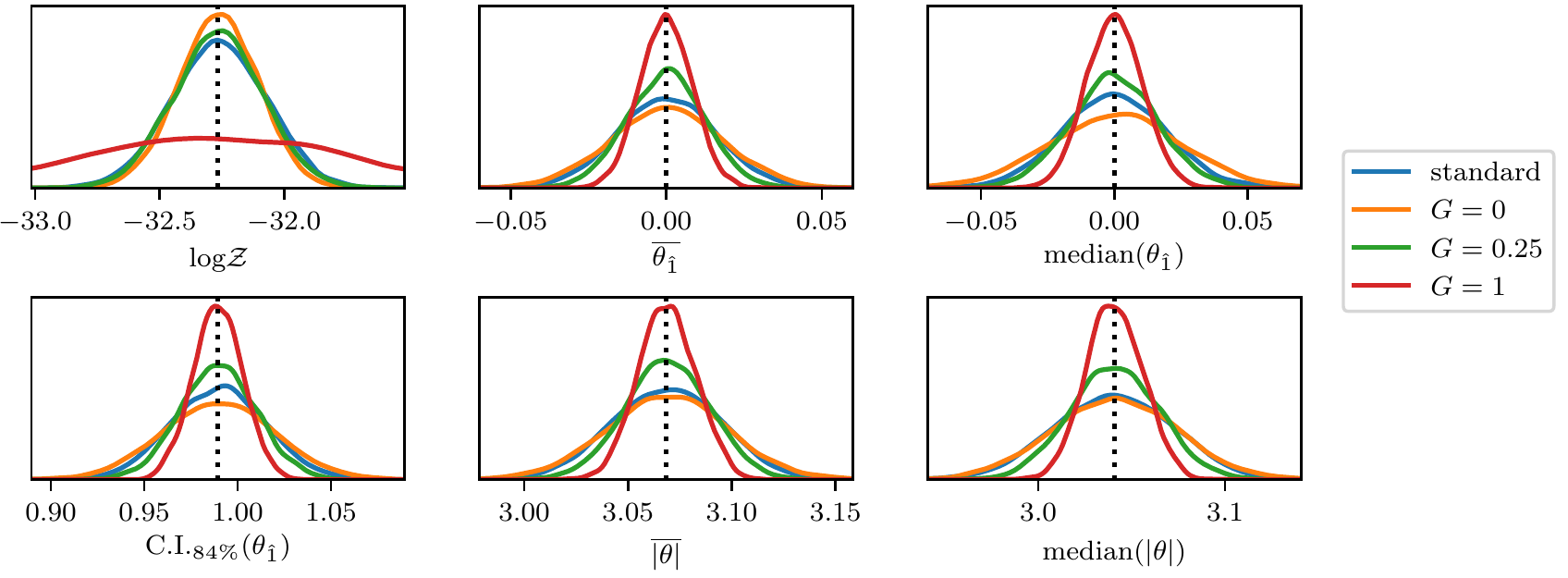}
    \caption{Distributions of results for the dynamic and standard nested sampling calculations shown in \Cref{tab:dynamic_test_gaussian}, plotted using kernel density estimation.
    Black dotted lines show the correct value of each quantity for the likelihood and prior used.
    Compared to standard nested sampling (blue lines), the distributions of results of dynamic nested sampling with $G=1$ (red lines) for parameter estimation problems show much less variation around the correct value.
    Results for dynamic nested sampling with $G=0$ (orange lines) are on average closer to the correct value than standard nested sampling for calculating $\log \mathcal{Z}$, and results with $G=0.25$ (green lines) show improvements over standard nested sampling for both evidence and parameter estimation calculations.
}\label{fig:dynamic_test_dists}
\end{figure*}

The variation of results from repeated standard and dynamic nested sampling calculations with a similar number of samples is shown in \Cref{tab:dynamic_test_gaussian} and \Cref{fig:dynamic_test_dists}.
Dynamic nested sampling optimised for evidence calculation ($G=0$) and parameter estimation ($G=1$) produce significantly more accurate results than standard nested sampling.
In addition, results for dynamic nested sampling with $G=0.25$ show that both evidence calculation and parameter estimation accuracy can be improved simultaneously.
Equivalent results for 10-dimensional exponential power likelihoods~\eqref{equ:exp_power} with $b=2$ and $b=\frac{3}{4}$ are shown in \Cref{tab:dynamic_test_exp_power_2,tab:dynamic_test_exp_power_0_75} in Appendix~\ref{app:exp_power_add_tests}.
The reduction in evidence errors for $G=0$ and parameter estimation errors for $G=1$ in \Cref{tab:dynamic_test_gaussian} correspond to increasing efficiency by factors of $1.40 \pm 0.04$ and up to $4.4 \pm 0.1$ respectively.

\begin{table*}
\centering
    \caption{Test of dynamic nested sampling for a 10-dimensional Gaussian likelihood~\eqref{equ:gaussian} and a Gaussian prior~\eqref{equ:gaussian_prior} with $\sigma_\pi = 10$.
The first row shows the standard deviation of $5,000$ calculations for standard nested sampling with a constant number of live points $n=500$.
The next three rows show the standard deviations of $5,000$ dynamic nested sampling calculations with a similar number of samples; these are respectively optimised purely for evidence calculation accuracy ($G=0$), for both evidence and parameter estimation ($G=0.25$) and purely for parameter estimation ($G=1$).
The final three rows show the computational efficiency gain~\eqref{equ:efficiency_gain} from dynamic nested sampling over standard nested sampling in each case.
The first column shows the mean number of samples for the $5,000$ runs.
The remaining columns show calculations of the log evidence, the mean, median and $84\%$ one-tailed credible interval of a parameter $\thcomp{1}$, and the mean and median of the radial coordinate $|\btheta|$.
Numbers in brackets show the $1\sigma$ numerical uncertainty on the final digit.\label{tab:dynamic_test_gaussian}}
\begin{tabular}{llllllll}
\toprule
{} &      samples & $\log \Z$ & $\po$ & $\mathrm{median}(\thcomp{1}) $ & $\mathrm{C.I.}_{84\%}(\thcomp{1})$ & $\overline{|\btheta|}$ & $\mathrm{median}(|\btheta|)$ \\
\midrule
St.Dev.\ standard             &  15,189 &                   0.189(2) &                     0.0158(2) &                           0.0194(2) &                                0.0253(3) &             0.0262(3) &                   0.0318(3) \\
St.Dev.\ $G=0$                &  15,152 &                   0.160(2) &                     0.0180(2) &                           0.0249(2) &                                0.0301(3) &             0.0292(3) &                   0.0335(3) \\
St.Dev.\ $G=0.25$             &  15,156 &                   0.179(2) &                     0.0124(1) &                           0.0163(2) &                                0.0204(2) &             0.0205(2) &                   0.0239(2) \\
St.Dev.\ $G=1$                &  15,161 &                   0.549(5) &                    0.00834(8) &                           0.0104(1) &                                0.0132(1) &             0.0138(1) &                   0.0152(2) \\
Efficiency gain $G=0$    &          &                    1.40(4) &                       0.77(2) &                             0.60(2) &                                  0.71(2) &               0.80(2) &                     0.90(3) \\
Efficiency gain $G=0.25$ &          &                    1.11(3) &                       1.62(5) &                             1.42(4) &                                  1.54(4) &               1.64(5) &                     1.77(5) \\
Efficiency gain $G=1$    &          &                   0.119(3) &                        3.6(1) &                              3.5(1) &                                   3.7(1) &                3.6(1) &                      4.4(1) \\
\bottomrule
\end{tabular}
\end{table*}

\subsection{Efficiency gains for different distributions of posterior mass}

Efficiency gains~\eqref{equ:efficiency_gain} from dynamic nested sampling depend on the fraction of the $\log X$ range explored which contains samples that make a significant contribution to calculation accuracy.
If this fraction is small most samples taken by standard nested sampling contain little information, and dynamic nested sampling can greatly improve performance.
For parameter estimation ($G=1$), only $\log X$ regions containing significant posterior mass ($\propto \mathcal{L}(X)X$) are important, whereas for evidence calculation ($G=0$) all samples taken before the bulk of the posterior is reached are valuable. Both cases benefit from dynamic nested sampling using fewer samples to explore the region after most of the posterior mass has been passed but before termination.

\begin{figure*}
\centering
\includegraphics{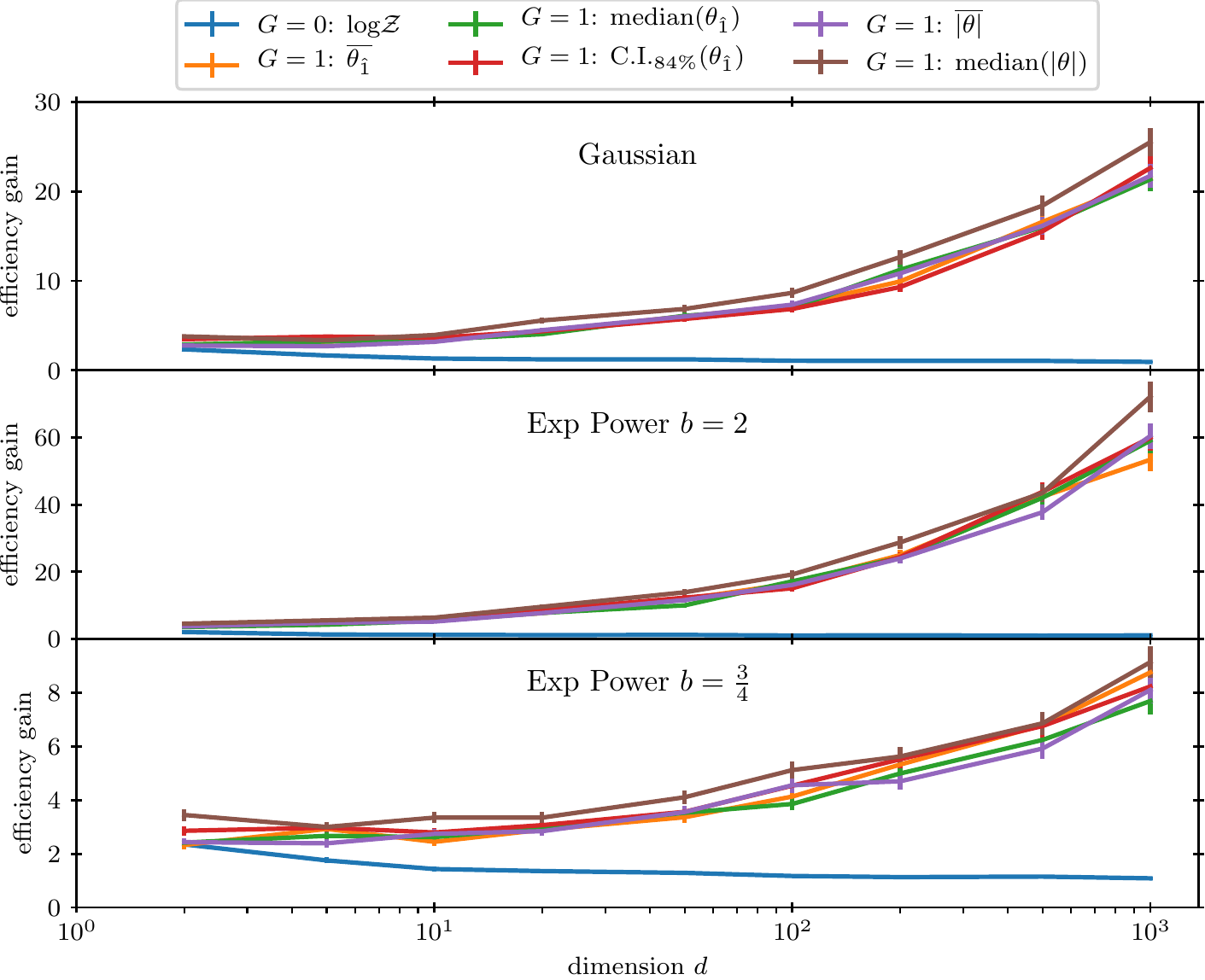}
\caption{Efficiency gain~\eqref{equ:efficiency_gain} from dynamic nested sampling compared to standard nested sampling for likelihoods of different dimensions; each has a Gaussian prior~\eqref{equ:gaussian_prior} with $\sigma_\pi = 10$.
Results are shown for calculations of the log evidence, the mean, median and $84\%$ one-tailed credible interval of a parameter $\thcomp{1}$, and the mean and median of the radial coordinate $|\btheta|$.
Each efficiency gain is calculated using $1,000$ standard nested sampling calculations with $n=200$ and $1,000$ dynamic nested sampling calculations with $\ninit=20$ using a similar or slightly smaller number of samples.}\label{fig:n_dim_performance}
\vspace{0.4cm}
\centering
\includegraphics{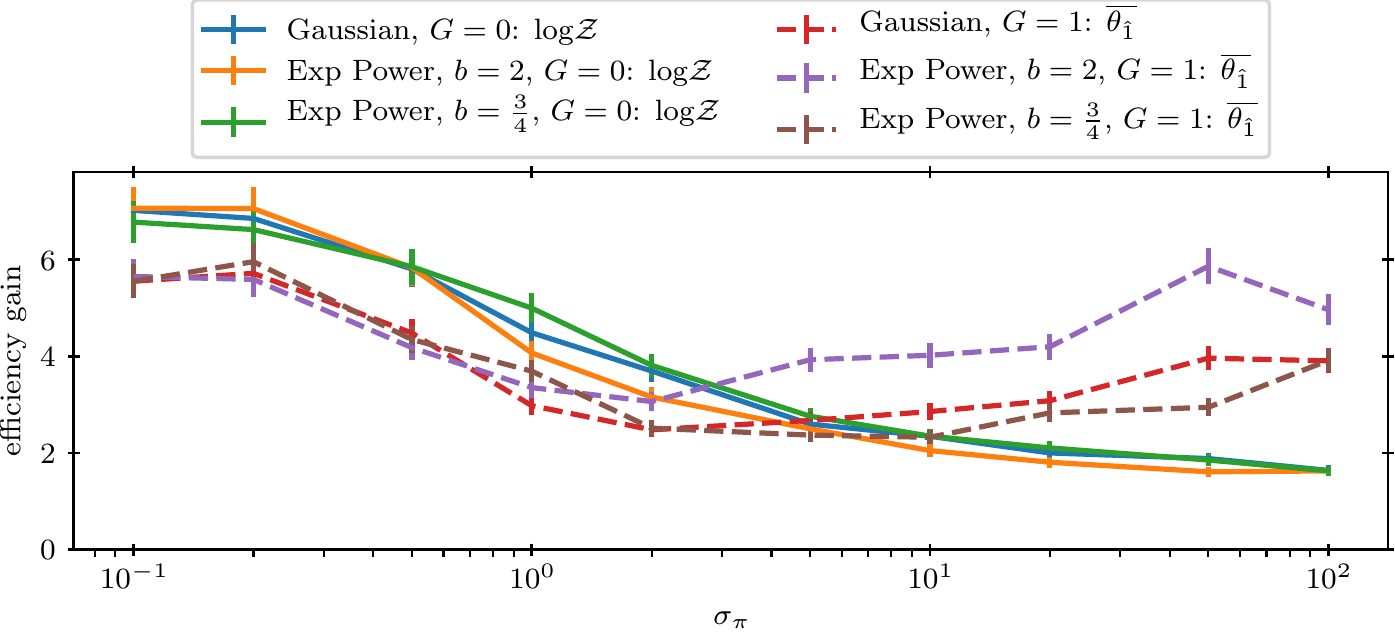}
\caption{Efficiency gain~\eqref{equ:efficiency_gain} from dynamic nested sampling for Gaussian priors~\eqref{equ:gaussian_prior} of different sizes $\sigma_\pi$.
Results are shown for calculations of the log evidence and the mean of a parameter $\thcomp{1}$ for $2$-dimensional Gaussian likelihoods~\eqref{equ:gaussian} and 2-dimensional exponential power likelihoods~\eqref{equ:exp_power} with $b=2$ and $b=\frac{3}{4}$.
    Each efficiency gain is calculated using $1,000$ standard nested sampling calculations with $n=200$ and $1,000$ dynamic nested sampling calculations with $\ninit=20$ using a similar or slightly smaller number of samples.}\label{fig:prior_r_max_performance}
\end{figure*}

We now test the efficiency gains~\eqref{equ:efficiency_gain} of dynamic nested sampling empirically for a wide range of distributions of posterior mass by considering Gaussian likelihoods~\eqref{equ:gaussian} and exponential power likelihoods~\eqref{equ:exp_power} of different dimensions $d$ and prior sizes $\sigma_\pi$.
The results are presented in \Cref{fig:prior_r_max_performance,fig:n_dim_performance}, and show large efficiency gains from dynamic nested sampling for parameter estimation in all of these cases.

Increasing the dimension $d$ typically means the posterior mass is contained in a smaller fraction of the prior volume \citep{Higson2017a}, as shown in \Cref{fig:an_w}.
In the spherically symmetric cases we consider, the range of $\log X$ to be explored before significant posterior mass is reached increases approximately linearly with $d$.
This increases the efficiency gain~\eqref{equ:efficiency_gain} from dynamic nested sampling for parameter estimation ($G=1$) but reduces it for evidence calculation ($G=0$).
In high-dimensional problems the vast majority of the $\log X$ range explored is usually covered before any significant posterior mass is reached, resulting in very large efficiency gains for parameter estimation but almost no gains for evidence calculation --- as can be seen in \Cref{fig:n_dim_performance}.
For the 1,000-dimensional exponential power likelihood with $b=2$, dynamic nested sampling with $G=1$ improves parameter estimation efficiency by a factor of up to $72\pm5$, with the largest improvement for estimates of the median the posterior distribution of $|\btheta|$.

Increasing the size of the prior $\sigma_\pi$ increases the fraction of the $\log X$ range explored before any significant posterior mass is reached, resulting in larger efficiency gains~\eqref{equ:efficiency_gain} from dynamic nested sampling for parameter estimation ($G=1$) but smaller gains for evidence calculation ($G=0$).
However when $\sigma_\pi$ is small the bulk of the posterior mass is reached after a small number of steps, and most of the $\log X$ range explored is after the majority of the posterior mass but before termination.
Dynamic nested sampling places fewer samples in this region than standard nested sampling, leading to large efficiency gains for both parameter estimation and evidence calculation.
This is shown in \Cref{fig:prior_r_max_performance}; when $\sigma_\pi = 0.1$, dynamic nested sampling evidence calculations with $G=0$ improve efficiency over standard nested sampling by a factor of approximately 7 for all 3 likelihoods considered.
However we note that if only the evidence estimate is of interest then standard nested sampling can safely terminate with a higher fraction of the posterior mass remaining than $10^{-3}$, in which case efficiency gains would be lower.

\section{Dynamic nested sampling with challenging posteriors}\label{sec:practical_problems}

Nested sampling software such as \MultiNest{} and \PolyChord{} use numerical techniques to perform the sampling within hard likelihood constrains required by the nested sampling algorithm; see \citet{Feroz2013,Handley2015b} for more details.
For challenging problems, such as those involving degenerate or multimodal posteriors, samples produced may not be drawn uniformly from the region of the prior within the desired iso-likelihood contour --- for example if this software misses a mode in a multimodal posterior.
This introduces additional uncertainties which are specific to a given software package and are not present in perfect nested sampling; we term these {\em implementation-specific effects\/} \citep[see][for a detailed discussion]{Higson2018a}.

\begin{figure}
	\centering
    \includegraphics[width=\linewidth]{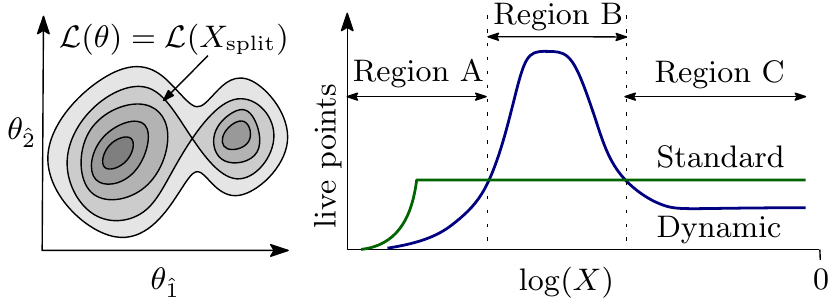}
    \caption{Dynamic and standard nested sampling's relative ability to discover hard to locate modes is determined by the number of live points present at the likelihood $\mathcal{L}(X_\mathrm{split})$ at which a mode splits from the remainder of the posterior (illustrated on the left).
In the schematic graph on the right we would expect dynamic nested sampling to be better at finding modes than standard nested sampling in region B (where it has a higher number of live points) but worse in regions A and C.}\label{fig:mode_splitting}
\end{figure}

Nested sampling software generally uses the population of dead and live points to sample within iso-likelihood contours, and so taking more samples in the region of an iso-likelihood contour will reduce the sampler's implementation-specific effects.
As a result dynamic nested sampling typically has smaller implementation-specific effects than standard nested sampling in the regions of the posterior where it has a higher number of live points, but conversely may perform worse in regions with fewer live points.
For highly multimodal or degenerate likelihoods it is important all modes or other regions of significant posterior mass are found by the sampler --- dynamic nested sampling performs better than standard nested sampling at finding hard to locate modes which become separated from the remainder of the posterior at likelihood values where it has more live points,%
\footnote{However, if a mode is only discovered late in the dynamic nested sampling process then it may still be under-sampled due to not being present in threads calculated before it was found.} as illustrated schematically in \Cref{fig:mode_splitting}.

Provided no significant modes are lost we expect dynamic nested sampling to have lower implementation-specific effects than standard nested sampling, as it has more live points --- and therefore lower implementation-specific effects --- in the regions which have the largest effect on calculation accuracy.
If modes separate at likelihood values where dynamic nested sampling assigns few samples, $\ninit$ must be made large enough to ensure no significant modes are lost.
For highly multimodal posteriors, a safe approach is to set $\ninit$ high enough to find all significant modes, in which case dynamic nested sampling will use the remaining computational budget to minimise calculation errors.
Even if, for example, half of the computational budget is used on the initial exploratory run, dynamic nested sampling will still achieve over half of the efficiency gain compared to standard nested sampling that it could with a very small $\ninit$.

The remainder of this section presents empirical tests of dynamic nested sampling for two challenging problems in which significant implementation-specific effects are present.
Additional examples of dynamic nested sampling's application to practical problems in scientific research can be found in \citet{Orazio2018}, \citet{Zucker2018}, \citet{Higson2018b} and \citet{Guillochon2018}.

\subsection{Numerical tests with a multimodal posterior}\label{sec:gaussian_mix}

\begin{figure}
	\centering
    \includegraphics{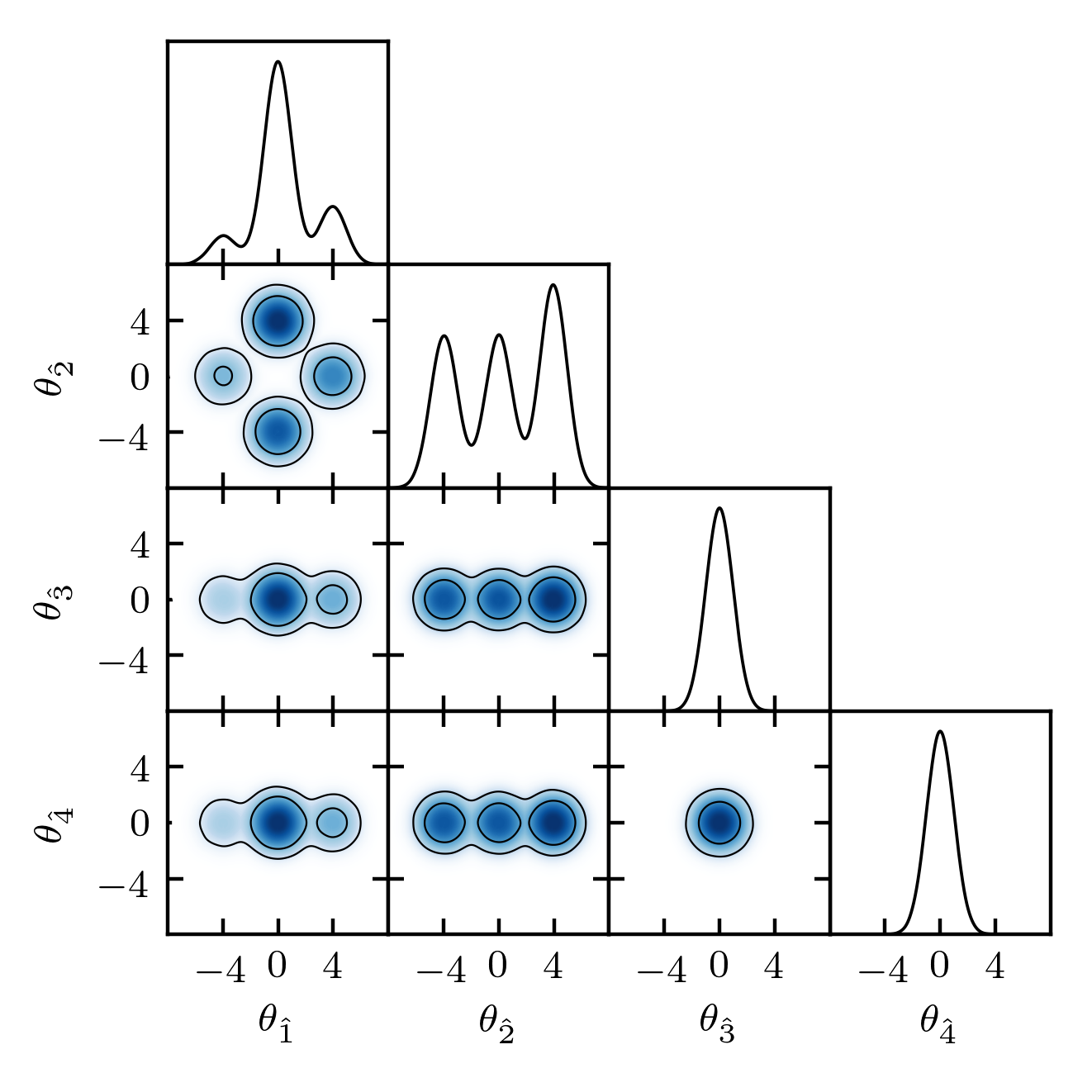}
    \caption{Posterior distributions for the 4-component 10-dimensional Gaussian mixture model~\eqref{equ:gaussian_mix} with component weights and means given by~\eqref{equ:mix_means}, and a Gaussian prior~\eqref{equ:gaussian_prior}.
By symmetry the distributions of $\thcomp{k}$ are the same for $k \in (3,\dots,d)$, so we only show only the first 4 components of $\btheta$; 1- and 2-dimensional plots of other parameters are the same as those of $\thcomp{3}$ and $\thcomp{4}$.}\label{fig:triangle_gaussian_mix}
\end{figure}

We now use \dyPolyChord{} to numerically test dynamic nested sampling on a challenging multimodal $d$-dimensional, $M$-component Gaussian mixture likelihood
\begin{equation}\label{equ:gaussian_mix}
    \mathcal{L}(\btheta) = \sum_{m=1}^M W^{(m)} {\left(2 \pi {\sigma^{(m)}}^2\right)}^{-d/2} \exp\left( -\frac{{|\btheta - \bmu^{(m)}|}^2}{2 {\sigma^{(m)}}^2}\right).
\end{equation}
Here each component $m$ is centred on a mean $\bmu^{(m)}$ with standard deviation $\sigma^{(m)}$ in all dimensions, and the component weights $W^{(m)}$ satisfy $\sum_{m=1}^M W^{(m)} = 1$.
For comparison with the perfect nested sampling results using a Gaussian likelihood~\eqref{equ:gaussian} in \Cref{sec:numerical_tests}, we use $d=10$, $\sigma^{(m)}=1$ for all $m$ and a Gaussian prior~\eqref{equ:gaussian_prior} with $\sigma_\pi = 10$.
We consider a Gaussian mixture~\eqref{equ:gaussian_mix} of $M=4$ components with means and weights
\begin{align}
    \quad W^{(1)}&=& 0.4, \qquad  \mu^{(1)}_{\hat{1}} &=&  0, \qquad \mu^{(1)}_{\hat{2}} &=& 4,\nonumber  \\
    \quad W^{(2)}&=& 0.3, \qquad  \mu^{(2)}_{\hat{1}} &=&  0, \qquad \mu^{(2)}_{\hat{2}} &=& -4,\nonumber \\
    \quad W^{(3)}&=& 0.2, \qquad  \mu^{(3)}_{\hat{1}} &=&  4, \qquad \mu^{(3)}_{\hat{2}} &=& 0,\label{equ:mix_means} \\
    \quad W^{(4)}&=& 0.1, \qquad  \mu^{(4)}_{\hat{1}} &=& -4, \qquad \mu^{(4)}_{\hat{2}} &=& 0,\nonumber
\end{align}
\vspace{-0.5cm}
\begin{equation*}
    \quad \text{and} \,\, \mu^{(m)}_{\hat{k}} = 0 \quad \text{for all} \,\, k  \, \in (3,\dots,d), \, m \in (1,\dots,M).
\end{equation*}
The posterior distribution for this case is shown in \Cref{fig:triangle_gaussian_mix}.

As in \Cref{sec:numerical_tests}, we compare standard nested sampling runs to dynamic nested sampling runs which use a similar or slightly smaller number of samples.
\dyPolyChord{} uses Algorithm~\ref{alg:dypolychord}, meaning only the initial run provides information on where to place samples, so we set $\ninit$ to 20\% of the number of live points used in standard nested sampling runs they are compared to, instead of the 10\% used in the perfect nested sampling tests in \Cref{sec:numerical_tests}.

\begin{figure*}
	\centering
    \includegraphics{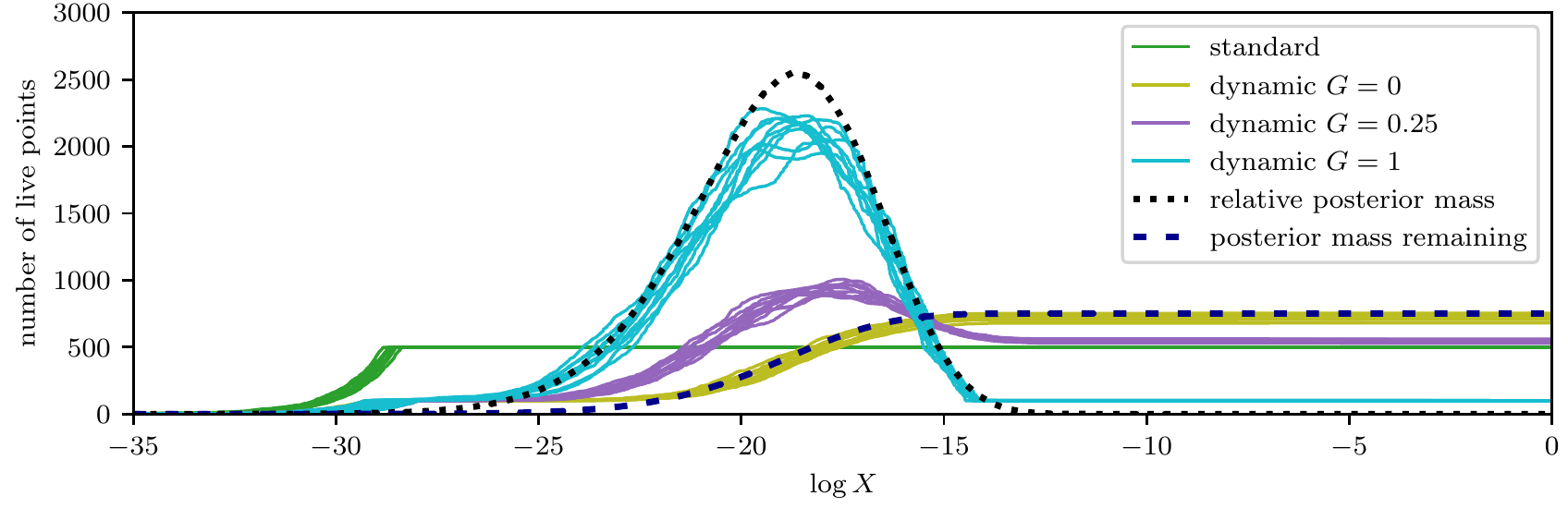}
    \caption{Live point allocation as in \Cref{fig:nlive_gaussian} but with a 10-dimensional Gaussian mixture likelihood~\eqref{equ:gaussian_mix}, with component weights and means given by~\eqref{equ:mix_means} and a Gaussian prior~\eqref{equ:gaussian_prior} with $\sigma_\pi = 10$.
    The 10 standard nested sampling runs shown were generated using \PolyChord{} with $n=500$, and 10 dynamic nested sampling runs with each $G$ value were generated using \dyPolyChord{} with a similar number of samples and $\ninit=100$.
    The dotted and dashed lines show the relative posterior mass $\propto \mathcal{L}(X)X$ and the posterior mass remaining $\propto \int_{-\infty}^X \mathcal{L}(X')X' \d{X'}$ at each point in $\log X$; for comparison these lines are scaled to have the same area under them as the average of the number of live point lines.
}\label{fig:nlive_gaussian_mix}
\end{figure*}

The allocation of live points from \dyPolyChord{} runs with the Gaussian mixture likelihood~\eqref{equ:gaussian_mix} is shown in~\Cref{fig:nlive_gaussian_mix}.
As in the tests with perfect nested sampling, the numbers of live points with settings $G=1$ and $G=0$ match the posterior mass and posterior mass remaining respectively despite the more challenging likelihood.
The live point allocation is not as precise as in~\Cref{fig:nlive_gaussian} due to \dyPolyChord{} only using information from the initial exploratory run to calculate all the point importances.
Another difference is that the truncation of the peak number of live points in the $G=1$ in \Cref{fig:nlive_gaussian} is not present for \dyPolyChord{} runs, as this is due to Algorithm~\ref{alg:dns} adding new points where the importance is within 90\% of the maximum.

\Cref{tab:dynamic_test_gaussian_mix} shows the variation of repeated calculations for dynamic nested sampling for the 10-dimensional Gaussian mixture model~\eqref{equ:gaussian_mix} with \dyPolyChord{}.
This shows significant efficiency gains~\eqref{equ:efficiency_gain} from dynamic nested sampling of $1.3 \pm 0.1$ for evidence calculation with $G=0$ and up to $4.0 \pm 0.4$ for parameter estimation with $G=1$, demonstrating how dynamic nested sampling can be readily applied to more challenging multimodal cases.
In Appendix~\ref{app:gaussian_mix_add_tests} we empirically verify that dynamic nested sampling does not introduce any errors from sampling bias (which would not be captured by efficiency gains~\eqref{equ:efficiency_gain} based on the variation of results) using analytically calculated true values of the log evidence and posterior means.
\Cref{tab:gaussian_mix_rmse} shows that the mean calculation results are very close to the correct values, and hence the standard deviation of the results is almost identical to their root-mean-squared-error, meaning efficiency gains~\eqref{equ:efficiency_gain} accurately reflect reductions in calculation errors (as for perfect nested sampling).

\begin{table*}
\centering
    \caption{Tests of dynamic nested sampling as in \Cref{tab:dynamic_test_gaussian} but with a 10-dimensional Gaussian mixture likelihood~\eqref{equ:gaussian_mix}, with component weights and means given by~\eqref{equ:mix_means} and a Gaussian prior~\eqref{equ:gaussian_prior} with $\sigma_\pi = 10$.
The first row shows the standard deviation of $500$ \PolyChord{} standard nested sampling calculations with a constant number of live points $n=500$.
The next three rows show the standard deviations of $500$ \dyPolyChord{} calculations with a similar number of samples; these are respectively optimised purely for evidence calculations ($G=0$), for both evidence and parameter estimation ($G=0.25$) and purely for parameter estimation ($G=1$).
All runs use the setting $\numrepeats=50$.
The final three rows show the computational efficiency gain~\eqref{equ:efficiency_gain} from dynamic nested sampling over standard nested sampling in each case.
The first column shows the mean number of samples produced by the $500$ runs.
The remaining columns show calculations of the log evidence, the mean of parameters $\thcomp{1}$ and $\thcomp{2}$, the median and $84\%$ one-tailed credible interval of $\thcomp{1}$, and the mean radial coordinate $|\btheta|$.
Numbers in brackets show the $1\sigma$ numerical uncertainty on the final digit.\label{tab:dynamic_test_gaussian_mix}}
\begin{tabular}{llllllll}
\toprule
{} & samples & $\log \mathcal{Z}$ & $\po$ & $\thmean{2}$ & $\mathrm{median}(\thcomp{1})$ & $\mathrm{C.I.}_{84\%}(\thcomp{1})$ & $\overline{|\btheta|}$ \\
\midrule
St.Dev.\ standard         &  14,739 &                   0.181(6) &                      0.057(2) &                      0.126(4) &                            0.035(1) &                                 0.170(5) &             0.0196(6) \\
St.Dev.\ $G=0$            &  14,574 &                   0.160(5) &                      0.076(2) &                      0.176(6) &                            0.048(2) &                                 0.229(7) &             0.0222(7) \\
St.Dev.\ $G=0.25$         &  14,628 &                   0.170(5) &                      0.046(1) &                      0.105(3) &                           0.0293(9) &                                 0.138(4) &             0.0156(5) \\
St.Dev.\ $G=1$            &  14,669 &                    0.36(1) &                      0.032(1) &                      0.069(2) &                           0.0203(6) &                                 0.085(3) &             0.0110(3) \\
Efficiency gain $G=0$    &     &                     1.3(1) &                       0.56(5) &                       0.51(5) &                             0.53(5) &                                  0.55(5) &               0.78(7) \\
Efficiency gain $G=0.25$ &     &                     1.1(1) &                        1.5(1) &                        1.5(1) &                              1.4(1) &                                   1.5(1) &                1.6(1) \\
Efficiency gain $G=1$    &     &                    0.25(2) &                        3.3(3) &                        3.4(3) &                              3.0(3) &                                   4.0(4) &                3.2(3) \\
\bottomrule
\end{tabular}
\end{table*}

\Cref{tab:imp_error_gaussian_mix} shows estimated implementation-specific effects for the results in \Cref{tab:dynamic_test_gaussian_mix}; these are calculated using the procedure described in \citet[][Section 5]{Higson2018a}, which estimates the part of the variation of results which is not explained by the intrinsic stochasticity of perfect nested sampling.
Dynamic nested sampling with $G=1$ and $G=0.25$ both reduce implementation-specific effects in all of the parameter estimation calculations as expected.
However we are not able to measure a statistically significant difference in implementation-specific effects for $\log \mathcal{Z}$ with $G=0$; this is because for evidence calculations implementation-specific effects represent a much smaller fraction of the total error \citep[see][for more details]{Higson2018a}.

\begin{table*}
\centering
\caption{Estimated errors due to implementation-specific effects for the Gaussian mixture likelihood results shown in \Cref{tab:dynamic_test_gaussian_mix}, calculated using the method described in \citet[][Section 5]{Higson2018a}.
Numbers in brackets show the $1\sigma$ numerical uncertainty on the final digit.}\label{tab:imp_error_gaussian_mix}
\begin{tabular}{lllllll}
\toprule
{} & $\log \mathcal{Z}$ & $\po$ & $\thmean{2}$ & $\mathrm{median}(\thcomp{1})$ & $\mathrm{C.I.}_{84\%}(\thcomp{1})$ & $\overline{|\btheta|}$ \\
\midrule
Implementation St.Dev.\ standard  &                    0.02(4) &                      0.044(2) &                      0.115(4) &                            0.022(2) &                                 0.138(7) &              0.005(3) \\
Implementation St.Dev.\  $G=0$    &                    0.06(2) &                      0.062(3) &                      0.163(6) &                            0.033(2) &                                 0.191(9) &              0.005(5) \\
Implementation St.Dev.\  $G=0.25$ &                    0.03(4) &                      0.035(2) &                      0.095(4) &                            0.018(2) &                                 0.110(6) &              0.002(4) \\
Implementation St.Dev.\  $G=1$    &                    0.00(8) &                      0.024(1) &                      0.062(2) &                            0.013(1) &                                 0.065(4) &              0.000(2) \\
\bottomrule
\end{tabular}
\end{table*}

The efficiency gains in \Cref{tab:dynamic_test_gaussian_mix} are slightly lower than those for the similar unimodal Gaussian likelihood~\eqref{equ:gaussian} used in~\Cref{tab:dynamic_test_gaussian}; this is because of the higher $\ninit$ value used, and because while implementation-specific effects are reduced by dynamic nested sampling they are not reduced by as large a factor as errors from the stochasticity of the nested sampling algorithm.

\subsection{Numerical tests with signal reconstruction from noisy data}\label{sec:fit}

We now test dynamic nested sampling on a challenging signal reconstruction likelihood, which fits a 1-dimensional function $y = f(x,\theta)$ using a sum of basis functions.
Similar signal reconstruction problems are common in scientific research and are of great practical importance; for a detailed discussion see \citet{Higson2018b}.

We consider reconstructing a signal $y(x)$ given $D$ data points $\{x_d,y_d\}$, each of which has independent Gaussian $x$- and $y$-errors of size $\sigma_x = \sigma_y = 0.05$ around their unknown true values $\{X_d,Y_d\}$.
In our example, the data points' true $x$-coordinates $X_d$ were randomly sampled with uniform probability in the range $0 < X_d < 1$.
In this case the likelihood is \citep{Hee2016a}
\begin{equation}
\begin{split}
    \mathcal{L}(\theta)
    =
    \prod_{d=1}^D \int_{0}^{1} \frac{\exp\left[-\frac{{(x_d-X_d)}^2}{2\sigma_x^2}-\frac{{(y_d-f(X_d,\theta))}^2}{2\sigma_y^2}\right]}{2\pi\sigma_x\sigma_y} \d{X_d},
    \label{equ:fitting_likelihood_hee}
\end{split}
\end{equation}
where the integrals are over the unknown true values of the data points' $x$-coordinates, and each likelihood calculation involves an integral for each of the $D$ data points.
We reconstruct the signal using generalised Gaussian basis functions
\begin{equation}
    \phi(x,a,\mu,\sigma,\beta) = a \e^{-{(|x - \mu|/\sigma)}^{\beta}},
    \label{equ:gg_1d}
\end{equation}
where when $\beta=1$ the basis function is proportional to a Gaussian.
Our reconstruction uses 4 such basis functions,\footnote{Here the number of basis functions used is fixed. Examples of signal reconstructions in which the number and form of the basis functions are determined from the data simultaneously can be found in \citet{Higson2018b}.} giving 16 parameters
\begin{equation}
   \theta=(a_1,a_2,a_3,a_4,\mu_1,\mu_2,\mu_3,\mu_4,\sigma_1,\sigma_2,\sigma_3,\sigma_4,\beta_1,\beta_2,\beta_3,\beta_4),
\end{equation}
and
\begin{equation}
    y(x,\theta) = \sum_{j=1}^4 \phi(x, a_j,\mu_j,\sigma_j,\beta_j).
\end{equation}
The priors used are given in \Cref{tab:fit_priors} in Appendix~\ref{app:fit}.

\begin{figure*}
	\centering
    \includegraphics{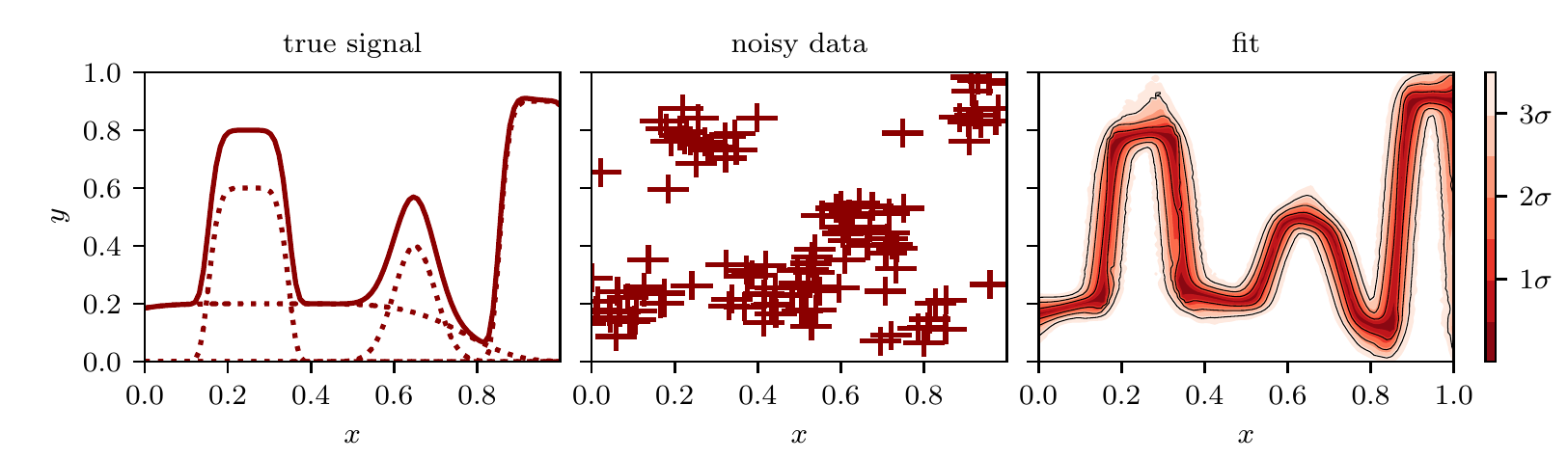}
    \caption{Signal reconstruction with generalised Gaussian basis functions.
    The first plot shows the true signal; this is composed of 4 generalised Gaussians~\eqref{equ:gg_1d}, with the individual components shown by dashed lines.
	The 120 data points, which have added normally distributed $x$- and $y$-errors with $\sigma_x=\sigma_y=0.05$, are show in the second plot.
    The third plot shows the fit calculated from a single \dyPolyChord{} dynamic nested sampling run with $G=1$, $\ninit=400$, $\numrepeats=400$ and 101,457 samples; coloured contours represent posterior iso-probability credible intervals on $y(x)$.}%
\label{fig:fit_fgivenx}
\end{figure*}

\begin{figure*}
	\centering
    \includegraphics{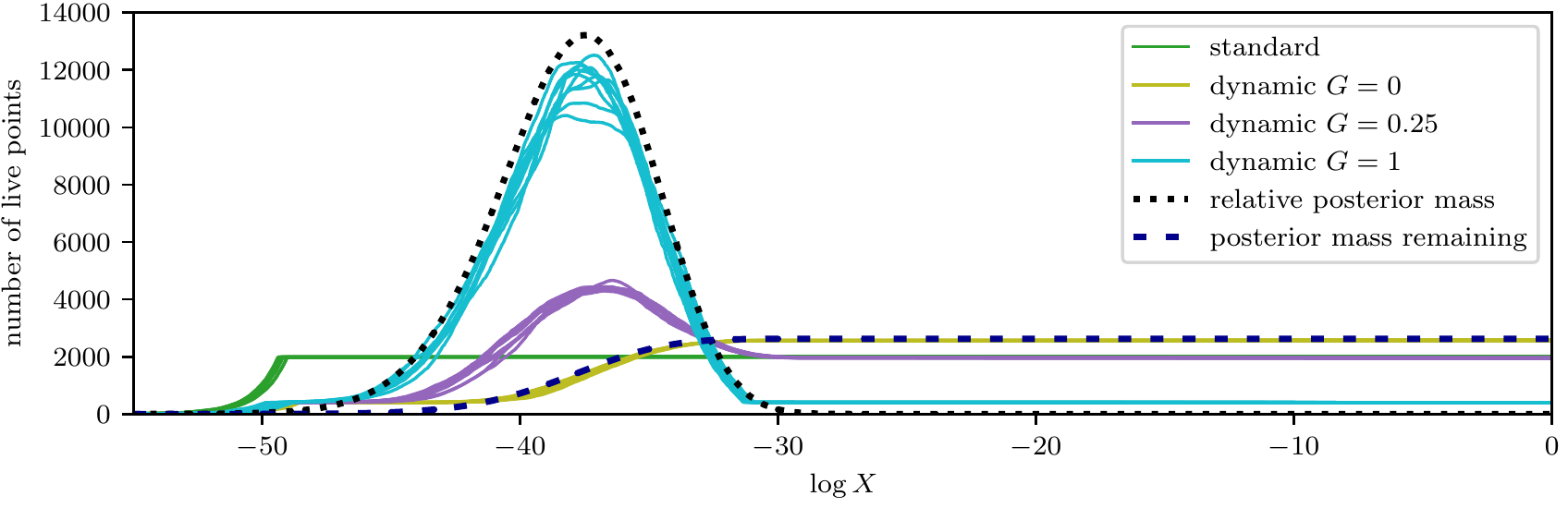}
    \caption{Live point allocation as in \Cref{fig:nlive_gaussian,fig:nlive_gaussian_mix} but for fitting 4 generalised Gaussians to the data shown in \Cref{fig:fit_fgivenx}.
    In this case the likelihood~\eqref{equ:fitting_likelihood_hee} is 16-dimensional, and the priors are given in \Cref{tab:fit_priors} in Appendix~\ref{app:fit}.
    The 10 standard nested sampling runs shown were generated using \PolyChord{} with $n=2,000$, and 10 dynamic nested sampling runs with each $G$ value were generated using \dyPolyChord{} with a similar number of samples and $\ninit=400$.
	All runs use the setting $\numrepeats=400$.
    The dotted and dashed lines show the relative posterior mass $\propto \mathcal{L}(X)X$ and the posterior mass remaining $\propto \int_{-\infty}^X \mathcal{L}(X')X' \d{X'}$ at each point in $\log X$; for comparison these lines are scaled to have the same area under them as the average of the number of live point lines.
}\label{fig:nlive_fit}
\end{figure*}

We use 120 data points, sampled from a true signal composed of the sum of 4 generalised Gaussian basis functions with parameters shown in \Cref{tab:fit_data_args} in Appendix~\ref{app:fit}.
The true signal, the noisy data and the posterior distribution of the signal calculated with dynamic nested sampling are shown in \Cref{fig:fit_fgivenx}; this was plotted using the \texttt{fgivenx} package \citep{Handley2018fgivenx}.
\dyPolyChord{}'s allocation of live points for the basis function fitting likelihood and priors are shown in \Cref{fig:nlive_fit}; as before, the software is able to accurately allocate live points in this case.

\Cref{tab:dynamic_test_fit} shows efficiency gains from dynamic nested sampling over standard nested sampling for the signal reconstruction problem.
Due to the computational expense of this likelihood, we use only 20 runs for each of standard nested sampling and dynamic nested sampling with $G=0$, $G=0.25$ and $G=1$.
Consequently the results are less precise than those for previous examples, but the improvements over standard nested sampling are similar to the other tests and include large efficiency gains in estimates of the mean value of the fitted signal (of up to $9.0\pm4.1$).
Furthermore, dynamic nested sampling is also able to reduce errors due to implementation-specific effects in this case --- as can be seen in \Cref{tab:imp_error_fit}.

\begin{table*}
\centering
\caption{Tests of dynamic nested sampling as in \Cref{tab:dynamic_test_gaussian,tab:dynamic_test_gaussian_mix} but for fitting 4 generalised Gaussians to the data shown in \Cref{fig:fit_fgivenx}; the likelihood is given by~\eqref{equ:fitting_likelihood_hee} and the priors are shown in \Cref{tab:fit_priors} in Appendix~\ref{app:fit}.
The first row shows the standard deviation of $20$ \PolyChord{} standard nested sampling calculations with a constant number of live points $n=2,000$.
The next three rows show the standard deviations of $20$ \dyPolyChord{} calculations with a similar number of samples; these are respectively optimised purely for evidence calculations ($G=0$), for both evidence and parameter estimation ($G=0.25$) and purely for parameter estimation ($G=1$).
The final three rows show the computational efficiency gain~\eqref{equ:efficiency_gain} from dynamic nested sampling over standard nested sampling in each case.
The first column shows the mean number of samples produced by the $20$ runs.
The remaining columns show calculations of the log evidence, and the posterior expectation of $y(x,\theta)$ at $x=0.1$, $x=0.3$, $x=0.5$, $x=0.7$ and $x=0.9$.
Numbers in brackets show the $1\sigma$ numerical uncertainty on the final digit.\label{tab:dynamic_test_fit}}
\begin{tabular}{llllllll}
\toprule
{}                        &  samples & $\log \mathcal{Z}$ & \ymean{0.1}& \ymean{0.3}& \ymean{0.5}& \ymean{0.7}& \ymean{0.9}\\
\midrule
St.Dev.\ standard         &  100,461 &            0.19(3) &  0.0013(2) &  0.0020(3) &  0.0020(3) &  0.0019(3) &  0.0016(3) \\
St.Dev.\ $G=0$            &  100,490 &            0.15(2) &  0.0018(3) &  0.0023(4) &  0.0025(4) &  0.0022(4) &  0.0027(4) \\
St.Dev.\ $G=0.25$         &  100,708 &            0.20(3) &  0.0015(2) &  0.0017(3) &  0.0018(3) &  0.0014(2) &  0.0017(3) \\
St.Dev.\ $G=1$            &  100,451 &            0.39(6) &  0.0007(1) &  0.0007(1) &  0.0009(1) &  0.0008(1) &  0.0013(2) \\
Efficiency gain $G=0$     &          &             1.7(8) &     0.5(2) &     0.8(3) &     0.7(3) &     0.7(3) &     0.4(2) \\
Efficiency gain $G=0.25$  &          &             0.9(4) &     0.8(4) &     1.4(7) &     1.2(6) &     1.7(8) &     1.0(4) \\
Efficiency gain $G=1$     &          &             0.2(1) &    3.6(16) &    9.0(41) &    4.9(22) &    5.2(24) &     1.7(8) \\
\bottomrule
\end{tabular}
\end{table*}

\begin{table*}
\centering
\caption{Estimated error due to implementation-specific effects for the basis function fitting likelihood results shown in \Cref{tab:dynamic_test_fit}, calculated using the method described in \citet[][Section 5]{Higson2018a}.
Numbers in brackets show the $1\sigma$ numerical uncertainty on the final digit.}\label{tab:imp_error_fit}
\begin{tabular}{lllllll}
\toprule
{} & $\log \mathcal{Z}$ & \ymean{0.1}& \ymean{0.3}& \ymean{0.5}& \ymean{0.7}& \ymean{0.9} \\
\midrule
Implementation St.Dev.\ standard  &                    0.14(5) &  0.0008(5) &  0.0016(4) &  0.0014(6) &  0.0013(6) &  0.0007(8) \\
Implementation St.Dev.\  $G=0$    &                    0.10(5) &  0.0012(6) &  0.0018(6) &  0.0017(7) &  0.0013(9) &  0.0021(6) \\
Implementation St.Dev.\  $G=0.25$ &                    0.16(5) &  0.0008(6) &  0.0010(6) &  0.0010(7) &  0.0000(7) &  0.0009(7) \\
Implementation St.Dev.\  $G=1$    &                     0.2(1) &  0.0000(3) &  0.0000(3) &  0.0000(4) &  0.0000(4) &  0.0010(3) \\
\bottomrule
\end{tabular}
\end{table*}

\section{Conclusion}

This paper began with an analysis of the effects of changing the number of live points on the accuracy of nested sampling parameter estimation and evidence calculations.
We then presented dynamic nested sampling (Algorithm~\ref{alg:dns}), which varies the number of live points to allocate posterior samples efficiently for {\em a priori\/} unknown likelihoods and priors.

Dynamic nested sampling can be optimised specifically for parameter estimation, showing increases in computational efficiency over standard nested sampling~\eqref{equ:efficiency_gain} by factors of up to $72\pm5$ in numerical tests.
The algorithm can also increase evidence calculation accuracy, and can improve both evidence calculation and parameter estimation simultaneously.
We discussed factors effecting the efficiency gain from dynamic nested sampling, including showing large improvements in parameter estimation are possible when the posterior mass is contained in a small region of the prior (as is typically the case in high-dimensional problems).
Empirical tests show significant efficiency gains from dynamic nested sampling for a wide range likelihoods, priors, dimensions and estimators considered.
Another advantage of dynamic nested sampling is that more accurate results can be obtained by continuing the run for longer, unlike in standard nested sampling.

We applied dynamic nested sampling to problems with challenging posteriors using \dyPolyChord{}, and found the technique is able to reduce errors due to implementation-specific effects compared to standard nested sampling.
This included tests with a practical signal reconstruction calculation, and a multimodal posterior in which the new method gave similar performance gains to the unimodal test cases.
Dynamic nested sampling has also been applied to a number of problems in scientific research; see for example \citet{Orazio2018}, \citet{Zucker2018}, \citet{Higson2018b} and \citet{Guillochon2018}.

The many popular approaches and software implementations for standard nested sampling can be easily adapted for dynamic nested sampling, since it too only requires samples to be drawn randomly from the prior within some hard likelihood constraint.
As a result, our new method can be used to increase computational efficiency while maintaining the strengths of standard nested sampling.
Publicly available dynamic nested sampling packages include \dyPolyChord{}, \dynesty{} and \perfectns{}.

\bibliographystyle{spbasic}      % basic style, author-year citations
\bibliography{library}
\begin{appendices}

\section*{Appendices}

\section{Code}

The code used to generate the numerical results and plots in this paper is available at \href{https://github.com/ejhigson/dns}{https://github.com/ejhigson/dns}.

\section{Estimating sampling errors in dynamic nested sampling}\label{app:bootstrap}

The technique for estimating sampling errors by resampling threads introduced in \citet{Higson2017a} can be applied to dynamic nested sampling runs with variable numbers of live points.
\Cref{tab:error_results} shows numerical tests of the bootstrap error estimates for dynamic nested sampling, calculated using the \nestcheck{} package \citep{Higson2018nestcheck}.
The results use $G=1$ --- this the most challenging case as most of the threads only cover part of the $\log X$ range explored by the run.
The bootstrap error estimates match the sampling errors observed when the calculation is repeated many times, in agreement with the results for standard nested sampling in \citet{Higson2017a}.

When $\ninit$ is low and $G=1$, bootstrap replications may contain zero (or very few) threads which begin by sampling the whole prior.
This typically does not matter for calculating parameter estimation errors as only the relative weights of points are used, but may lead to inaccurate estimates of evidence errors.
In this case the threads from the initial exploratory run can be sampled separately (with replacement), ensuring every bootstrap replication contains $\ninit$ such threads --- this approach was used for~\Cref{tab:error_results}.
When $\ninit$ is close to 1, estimates of $\log \mathcal{Z}$ uncertainties with this approach become imprecise, and the simulated weights method may perform better \citep[see][for more details]{Higson2017a}.

\begin{table*}
    \caption{Bootstrap sampling error estimates for dynamic nested sampling of a 3-dimensional Gaussian likelihood~\eqref{equ:gaussian} and a Gaussian prior~\eqref{equ:gaussian_prior}
The table shows results from 5,000 dynamic nested sampling runs generated with \perfectns{} using $G=1$, $\ninit=20$ and with the same total number of samples as standard nested sampling with a constant $n=200$ live points.
The first two rows show the mean and standard deviation of the results of the $5,000$ calculations.
The third row shows the mean of the error estimates from the bootstrap resampling technique for each run (using $200$ replications), divided by the error observed from repeated calculations.
The fourth row shows the standard deviations of bootstrap error estimates for single runs as a percentage of the mean estimate.
The fifth row shows the mean of $500$ bootstrap estimates of the one-tailed $95\%$ credible interval on the calculation result given the sampling error, each using $1,000$ bootstrap replications.
The final two rows show the empirical coverage of the bootstrap standard error and $95\%$ credible interval from the $5,000$ repeated calculations.
Numbers in brackets show the $1\sigma$ numerical uncertainty on the final digit.\label{tab:error_results}}
\centering
\begin{tabular}{lllllll}
\toprule
{} & $\log \mathcal{Z}$ & $\po$ & $\mathrm{median}(\thcomp{1})$ & $\mathrm{C.I.}_{84\%}(\thcomp{1})$ & $\overline{|\btheta|}$ & $\mathrm{median}(|\btheta|)$ \\
\midrule
Mean result                                 &                  -9.710(7)   &                     0.0002(3)    &                           0.0003(3)    &                                0.9904(4)    &             1.5890(3)    &                   1.5316(3)    \\
Repeated runs St.Dev.\                        &                   0.464(5)   &                     0.0184(2)    &                           0.0234(2)    &                                0.0294(3)    &             0.0195(2)    &                   0.0232(2)    \\
Bootstrap St.Dev.\ / Repeats St.Dev.\           &                    0.99(1)   &                       1.02(1)    &                             1.00(1)    &                                  1.03(1)    &               1.01(1)    &                     1.00(1)    \\
Bootstrap $\std$ estimate variation         &                    17.1(2)\% &                       6.07(6)\%  &                             11.5(1)\%  &                                  13.1(1)\%  &               6.69(7)\%  &                     10.9(1)\%  \\
Bootstrap $\mathrm{C.I.}_{95\%}$            &                   -8.94(2)   &                     0.0304(8)    &                            0.038(1)    &                                 1.038(1)    &             1.6209(9)    &                    1.569(1)    \\
Bootstrap Mean$\pm1\std$ coverage           &                       67.7\% &                          68.6\%  &                                68.4\%  &                                       70\%  &                  68.5\%  &                        69.0\%  \\
Bootstrap $\mathrm{C.I.}_{95\%}$ coverage   &                       95.6\% &                          94.9\%  &                                94.7\%  &                                     95.0\%  &                  95.2\%  &                        94.8\%  \\
\bottomrule
\end{tabular}
\end{table*}

\section{Effect of varying the number of live points on evidence calculation accuracy}\label{app:optimum_z_derivation}

Nested sampling estimates the Bayesian evidence $\mathcal{Z}$ as the expectation of~\eqref{equ:ztot}, as described in~\Cref{sec:background}.
The dominant source of uncertainty is the unknown shrinkage ratios $t_i$, which are independent random variables with probability density functions $P(t_i)$ given in~\eqref{equ:dist_t}.
We now investigate the effect of increasing the number of live points $n_i$ across some shrinkage $t_i$ by considering~\eqref{equ:ztot} with all $t_{j\ne i}$ marginalised out and conditioned on $t_i$, defining
\begin{equation}
    \mathcal{Z}(t_i) \equiv \int \left( \sum_j w_j(\mathbf{t}) \mathcal{L}_j \right) \prod_{j\ne i} P(t_j) \d{t_j}.\label{equ:deff_z_t_i}
\end{equation}
For brevity in the remainder of this section we omit the explicit dependence of quantities such as point weights $w_i(\mathbf{t})$ on the shrinkage ratios $\mathbf{t}$.

For simplicity instead of using the trapezium rule we calculate point weight as
\begin{equation}
	w_i = X_{i-1} - X_{i} = (1 - t_i) \prod_{k<i} t_k.\label{equ:weights_trap}
\end{equation}
In this case uncertainty in $t_i$ causes sampling errors in the weight of point $i$ and all subsequent points\footnote{If the trapezium rule is used $t_i$ also affects the weight of the previous point $i-1$, but this has little effect on the results.} and
\begin{equation}
    \begin{split}
    \sum_j w_j \mathcal{L}_j
    =
    &\left[ \sum_{j<i} w_j \mathcal{L}_j \right] + (1 - t_i) \left[ \frac{w_i \mathcal{L}_i}{1 - t_i} \right]
    \\
    &+ t_i \left[ \sum_{j>i} \frac{w_j \mathcal{L}_j}{t_i} \right],
    \end{split}
\label{equ:rewrite_sum}
\end{equation}
where the terms in square brackets are independent of $t_i$. Substituting~\eqref{equ:rewrite_sum} into~\eqref{equ:deff_z_t_i} and integrating gives
\begin{equation}
    \mathcal{Z}(t_i) = \mathrm{E}[\mathcal{Z}_{<i}] + (1-t_i)\mathrm{E}\left[ \frac{\mathcal{L}_i w_i}{1 - t_i} \right] + t_i \mathrm{E}\left[\frac{\mathcal{Z}_{>i}}{t_i}\right],\label{equ:zti_unsimp}
\end{equation}
where we have defined $\mathcal{Z}_{>i} \equiv \sum_{k>i} \mathcal{L}_k w_k$ and $\mathcal{Z}_{<i} \equiv \sum_{k<i} \mathcal{L}_k w_k$.
The second term can be simplified by observing that as the shrinkage ratios are independent $\mathcal{L}_i w_i / (1 - t_i)$ is uncorrelated with $(1 - t_i)$, and that from~\eqref{equ:weights_trap} $\mathcal{L}_i w_i \propto (1 - t_i)$.
Two uncorrelated random variables $A$ and $B$ must satisfy $\mathrm{E}[A] = \mathrm{E}[AB]/\mathrm{E}[B]$, so hence
\begin{equation}
 \mathrm{E}\left[ \frac{\mathcal{L}_i w_i}{1 - t_i} \right]
 =
 \frac{\mathrm{E}[\mathcal{L}_i w_i]}{\mathrm{E}[1 - t_i]}
 =
 \frac{\mathrm{E}[\mathcal{L}_i w_i]}{1 - \mathrm{E}[t_i]}.
\end{equation}
Similarly $\mathcal{Z}_{>i} / t_i$ is uncorrelated with $t_i$ and from~\eqref{equ:weights_trap} $\mathcal{Z}_{>i} \propto t_i$, so
\begin{equation}
    \mathrm{E}\left[\frac{\mathcal{Z}_{>i}}{t_i}\right] = \frac{\mathrm{E}[\mathcal{Z}_{>i}]}{\mathrm{E}[t_i]}.
\end{equation}
Hence~\eqref{equ:zti_unsimp} can be rewritten as 
\begin{equation}
    \mathcal{Z}(t_i) = \mathrm{E}[\mathcal{Z}_{<i}] + (1-t_i)\frac{\mathrm{E}[\mathcal{L}_i w_i]}{(1 - \mathrm{E}[t_i])} + t_i \frac{\mathrm{E}[\mathcal{Z}_{>i}]}{\mathrm{E}[t_i]}.\label{equ:zti}
\end{equation}
Furthermore, from the distribution of the shrinkage ratios~\eqref{equ:dist_t}
\begin{equation}
    \mathrm{E}[t_i]        = \frac{n_i}{1+n_i},   \qquad
    \std[t_i]    = \frac{{n_i}^{1/2}}{(n_i+1){(n_i+2)}^{1/2}}.
    \label{equ:dist_t2}
\end{equation}
Substituting this into~\eqref{equ:zti} gives
\begin{equation}
    \begin{split}
    \mathcal{Z}(t_i) = &\bigg( \mathrm{E}[\mathcal{Z}_{<i}] + \mathrm{E}[\mathcal{L}_i w_i](n_i+1) \bigg)
    \\
    & + t_i\bigg(\frac{n_i+1}{n_i}\mathrm{E}[\mathcal{Z}_{>i}] - (1+n_i)\mathrm{E}[\mathcal{L}_i w_i]\bigg),
    %using \bigg( not \left( as \left( has autoscaling and is normal size without fraction
    \end{split}
\end{equation}
where terms in large brackets are independent of $t_i$. Using the expression for $\std[t_i]$ from~\eqref{equ:dist_t2}, the standard deviation of $\mathcal{Z}(t_i)$ is
\begin{equation}
    \begin{split}
    \std[\mathcal{Z}(t_i)]
    =&
    \frac{1}{{n_i}^{1/2} {(n_i+2)}^{1/2}}  \mathrm{E}[\mathcal{Z}_{>i}]
    \\
    &- \frac{n_i^{1/2}}{{(n_i+2)}^{1/2}} \mathrm{E}[\mathcal{L}_i w_i].
    \end{split}
\end{equation}
The expected number of samples (computational work) needed to increase the number of live points over some interval $(\mathcal{L}_a, \mathcal{L}_b)$ is proportional to the log prior shrinkage $\log X (\mathcal{L}_a) - \log X (\mathcal{L}_b)$.
Hence the expected extra samples $\Delta N_\mathrm{s}$ required to increase the local number of live points $n_i$ is proportional to the interval $\log t_i$, which has an expected size of $1/n_i$.
The change in the error on the evidence with extra samples is therefore
\begin{eqnarray}
    \frac{\d{}}{\d{N_\mathrm{s}}}\std[\mathcal{Z}(t_i)] &=& \frac{\d{n_i}}{\d{N_\mathrm{s}}} \frac{\d{}}{\d{n_i}} \std[\mathcal{Z}(t_i)]\\
    & \propto &
    n_i \, \frac{\d{}}{\d{n_i}} \std[\mathcal{Z}(t_i)]\\
    & \propto &
    \label{equ:exact_z_importance}
    - \frac{n_i + 1}{{n_i}^{1/2} {(n_i+2)}^{3/2} } \mathrm{E}[\mathcal{Z}_{>i}]
    \\
    &&- \frac{n_i^{1/2}}{{(n_i+2)}^{3/2}} \mathrm{E}[\mathcal{L}_i w_i].\nonumber
\end{eqnarray}
This quantity can be easily calculated for a set of dead points with little computational cost.
Typically $n_i \gg 2$, in which case the following relation approximately holds:
\begin{equation}
    \frac{\d{}}{\d{N_\mathrm{s}}}\std[\mathcal{Z}(t_i)] \propto - \frac{\mathrm{E}[\mathcal{Z}_{\ge i}]}{n_i},
\end{equation}
where $\mathcal{Z}_{\ge i} \equiv \sum_{k \ge i} \mathcal{L}_k w_k(\mathbf{t})$.
Thus the accuracy gained from taking additional samples is approximately proportional to the evidence contained in subsequent dead points.
This makes sense as the dominant evidence errors are from statistically estimating shrinkages $t_i$ which affect all subsequent points $j\ge i$.

\section{Tuning for a specific parameter estimation problem}\label{app:tuning}

Dynamic nested sampling improves parameter estimation efficiency by placing more samples in $\log X$ regions with significant posterior mass and fewer in regions with little posterior mass.
However, for some likelihoods and parameter estimation problems a large contribution to errors comes from samples in $\log X$ regions containing extreme or highly variable parameter values but little posterior weight (see Section 3.1 of \citet{Higson2017a} for a diagrammatic illustration).
In this case the expression for sample importances~\eqref{equ:p_importance} can be modified to favour points with parameter values which will have a large effect on the calculation.

For example, when estimating the global mean of some parameter or function of parameters $\mathrm{E}[f(\btheta)] = \sum_i f(\btheta_i) \mathcal{L}_i w_i$, one could place additional weight on regions with parameter values that have a large effect on results by calculating importances as
\begin{equation}
    \importancep(i) \propto \left| f(\btheta_i) - \mathrm{E}[f(\btheta)] \right| \mathcal{L}_i w_i.\label{equ:tuned_mean_general}
\end{equation}
This expression is highly variable as each point $i$ is a single sample from an iso-likelihood contour $\mathcal{L}(\btheta) = \mathcal{L}_i$ which may cover a wide range of parameters.
However dynamic nested sampling (Algorithm~\ref{alg:dns}) uses only the first and last points of high importance in allocating new threads, so~\eqref{equ:tuned_mean_general} captures $\log X$ regions in which some samples have extreme or highly variable parameter values.
When tuning dynamic nested sampling for calculating the mean of a parameter $\thcomp{1}$,~\eqref{equ:tuned_mean_general} becomes
\begin{equation}
    \importancep(i) \propto \left| \theta_{i,\hat{1}} - \po \right| \mathcal{L}_i w_i, \label{equ:tuned_importance}
\end{equation}
where $\po$ is the global mean of $\thcomp{1}$ and $\theta_{i,\hat{1}}$ is the $i$\textsuperscript{th} sample's $\thcomp{1}$ value.

\begin{figure*}
	\centering
    \includegraphics{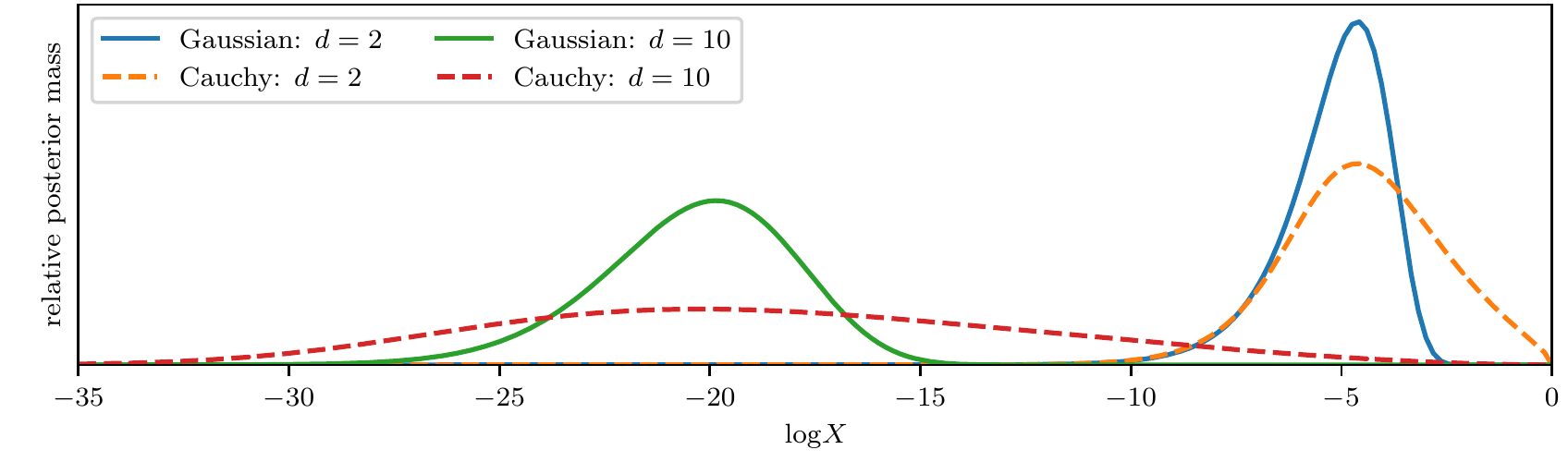}
    \caption{Relative posterior mass ($\propto \mathcal{L}(X)X$) as a function of $\log X$ for Cauchy likelihoods~\eqref{equ:cauchy}, with Gaussian likelihoods~\eqref{equ:gaussian} shown for comparison. Each has a Gaussian prior~\eqref{equ:gaussian_prior} with $\sigma_\pi=10$.
The lines are scaled so that the area under each of them is equal.}\label{fig:an_w_cauchy}
\end{figure*}

We illustrate tuning for a specific parameter by using dynamic nested sampling with a $d$-dimensional spherical unit Cauchy likelihood
\begin{equation}\label{equ:cauchy}
    \mathcal{L}(\btheta)=\frac{\Gamma(\frac{1+d}{2})}{\pi^{(d+1)/2}}{\left(1+{|\btheta|}^2\right)}^{-(\frac{d+1}{2})}.
\end{equation}
The Cauchy likelihoods have extremely heavy tails and (except in high dimensions) have significant posterior mass present across almost the entire range of $\log X$ explored, as shown in \Cref{fig:an_w_cauchy}.
We therefore expect relatively low efficiency gains for dynamic parameter estimation ($G=1$) in this case, but use it for a proof of principle.

\begin{figure*}
	\centering
    \includegraphics{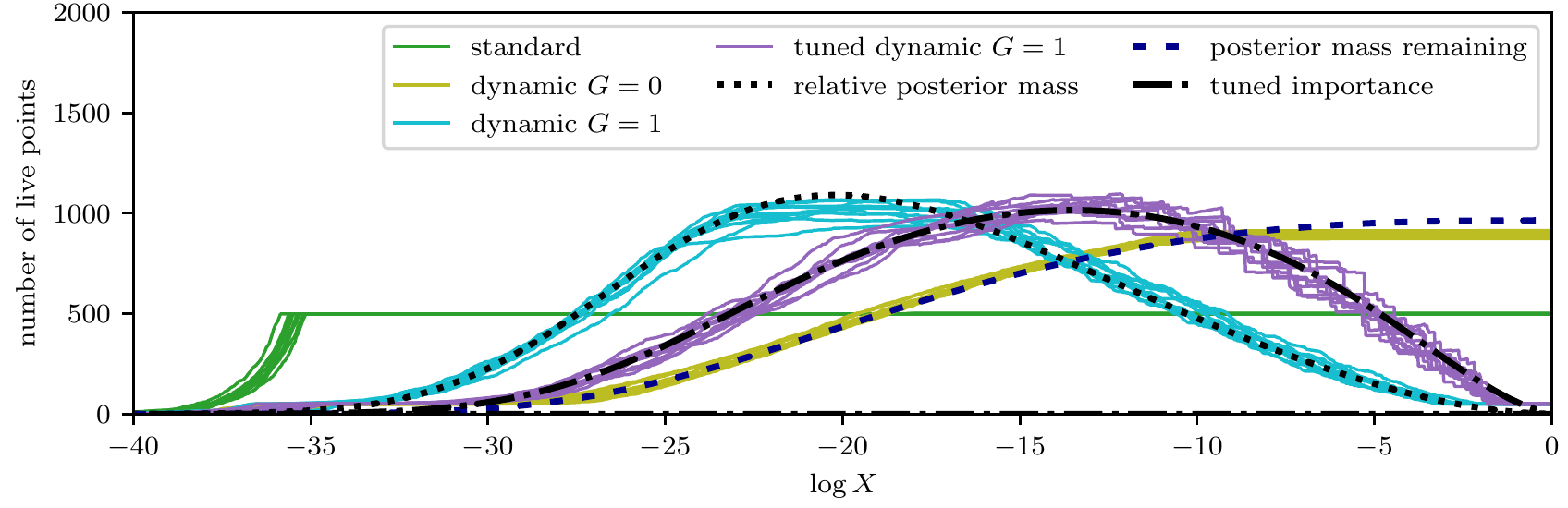}
    \caption{Live point allocation for a 10-dimensional Cauchy likelihood~\eqref{equ:cauchy} with a Gaussian prior~\eqref{equ:gaussian_prior} and $\sigma_\pi = 10$.
    Solid green lines show the number of live points as a function of $\log X$ for 10 standard nested sampling runs.
    Solid yellow, blue and purple lines show 10 dynamic nested sampling runs with $G=0$, $G=1$ and $G=1$ with a tuned importance function~\eqref{equ:tuned_importance} respectively.
    Dynamic runs use a similar number of samples to standard runs.
The dotted, dashed and dot-and-dash lines show the relative posterior mass $\propto \mathcal{L}(X)X$, the posterior mass remaining $\propto \int_{-\infty}^X \mathcal{L}(X')X' \d{X'}$ and the analytical expectation of the tuned importance function~\eqref{equ:tuned_importance_analyt}.
For comparison these lines are scaled to have the same area under them as the average of the number of live point lines.\label{fig:nlive_tuned}}
\end{figure*}

For a Cauchy likelihood~\eqref{equ:cauchy} with a co-centred spherically symmetric uniform prior, the analytic value of $\mathrm{E}[\thcomp{1}]$ is 0 and each iso-likelihood contour $\mathcal{L}(\btheta)=\mathcal{L}(X)$ is a spherically symmetric surface with radius $|\btheta|$.
The expectation of $|\thcomp{i}|$ on such an iso-likelihood contour is $|\btheta|/\sqrt{d}$, so the analytical expectation of the importance~\eqref{equ:tuned_importance} is
\begin{equation}
    \importancep(X) \propto |\btheta|  X \mathcal{L}(X) / \sqrt{d}.\label{equ:tuned_importance_analyt}
\end{equation}
\Cref{fig:nlive_tuned} shows the allocation of live points by dynamic nested sampling with and without tuning.
The numbers of live points as a function of $\log X$ for the tuned runs are consistent with~\eqref{equ:tuned_importance_analyt}, showing that samples can be allocated accurately when the tuned importance function is used.

\Cref{tab:tuned_dynamic_test} shows the efficiency gain for dynamic nested sampling for a 10-dimensional Cauchy likelihood~\eqref{equ:cauchy} with a Gaussian prior~\eqref{equ:gaussian_prior} and $\sigma_\pi=10$.
When estimating $\po$ the calculation is dominated by samples in the tails of the distribution with low likelihoods.
As a result, compared to standard nested sampling, dynamic nested sampling with $G=1$ slightly increases the variation of results --- giving an efficiency gain~\eqref{equ:efficiency_gain} of less than 1.
Tuned dynamic nested sampling is able to improve the efficiency gain for $\po$, as shown in the final row of \Cref{tab:tuned_dynamic_test}, although for the Cauchy likelihood the resulting gain is still small.
Using the tuned importance function affects the performance gain for other quantities --- for example in this case it significantly improves estimates of the second moment of the distribution $\overline{\thcomp{1}^2}$ in comparison to the $G=1$ case without tuning, but reduces the accuracy of estimates of the 84\% credible interval of $\thcomp{1}$.

\begin{table*}
\centering
    \caption{Test of tuned dynamic nested sampling with a 10-dimensional Cauchy likelihood~\eqref{equ:cauchy}, and a Gaussian prior~\eqref{equ:gaussian_prior} with $\sigma_\pi = 10$.
The first four rows show the standard deviation of $1,000$ calculations for standard nested sampling and dynamic nested sampling with $G=0$, $G=1$ and with a tuned importance function~\eqref{equ:tuned_importance} and $G=1$.
The final three rows show the computational efficiency gain~\eqref{equ:efficiency_gain} from dynamic nested sampling over standard nested sampling in each case.
The first column shows the mean number of samples for the $1,000$ runs.
The remaining columns show calculations of the log evidence, the mean, second moment and $84\%$ one-tailed credible interval of the parameter $\thcomp{1}$, and the mean and median radial coordinate $|\btheta|$.
Numbers in brackets show the $1\sigma$ numerical uncertainty on the final digit.\label{tab:tuned_dynamic_test}}
\begin{tabular}{llllllll}
\toprule
{} &      samples & $\log \mathcal{Z}$ & $\po$ & $\overline{\thcomp{1}^2}$ & $\mathrm{C.I.}_{84\%}(\thcomp{1})$ & $\overline{|\btheta|}$ & $\mathrm{median}(|\btheta|)$ \\
\midrule
St.Dev.\ standard                &  18,209 &                   0.167(4) &                     0.0124(3) &                        0.238(5) &                                 0.055(1) &              0.180(4) &                    0.165(4) \\
St.Dev.\ $G=0$                   &  18,165 &                   0.133(3) &                     0.0119(3) &                        0.214(5) &                                 0.056(1) &              0.173(4) &                    0.165(4) \\
St.Dev.\ $G=1$                   &  18,181 &                   0.320(7) &                     0.0128(3) &                        0.236(5) &                                 0.044(1) &              0.157(4) &                    0.125(3) \\
St.Dev.\ $G=1$ tuned             &  18,181 &                   0.244(5) &                     0.0106(2) &                        0.185(4) &                                 0.045(1) &              0.141(3) &                    0.130(3) \\
Efficiency gain $G=0$       &          &                     1.6(1) &                       1.08(7) &                         1.23(8) &                                  0.97(6) &               1.08(7) &                     0.99(6) \\
Efficiency gain $G=1$       &          &                    0.27(2) &                       0.94(6) &                         1.01(6) &                                   1.6(1) &               1.32(8) &                      1.7(1) \\
Efficiency gain $G=1$ tuned &          &                    0.46(3) &                       1.35(9) &                          1.6(1) &                                   1.5(1) &                1.6(1) &                      1.6(1) \\
\bottomrule
\end{tabular}
\end{table*}

\section{Additional numerical tests}

\subsection{Exponential power likelihoods}\label{app:exp_power_add_tests}

This section contains additional tests of dynamic nested sampling using 10-dimensional exponential power likelihoods~\eqref{equ:exp_power} with $b=2$ and $b=\frac{3}{4}$; compared to Gaussian likelihoods~\eqref{equ:gaussian} these have lighter and heavier tails respectively.
As in \Cref{sec:numerical_tests}, each test uses a Gaussian prior~\eqref{equ:gaussian} with $\sigma_\pi =10$.

\begin{figure*}
	\centering
    \includegraphics{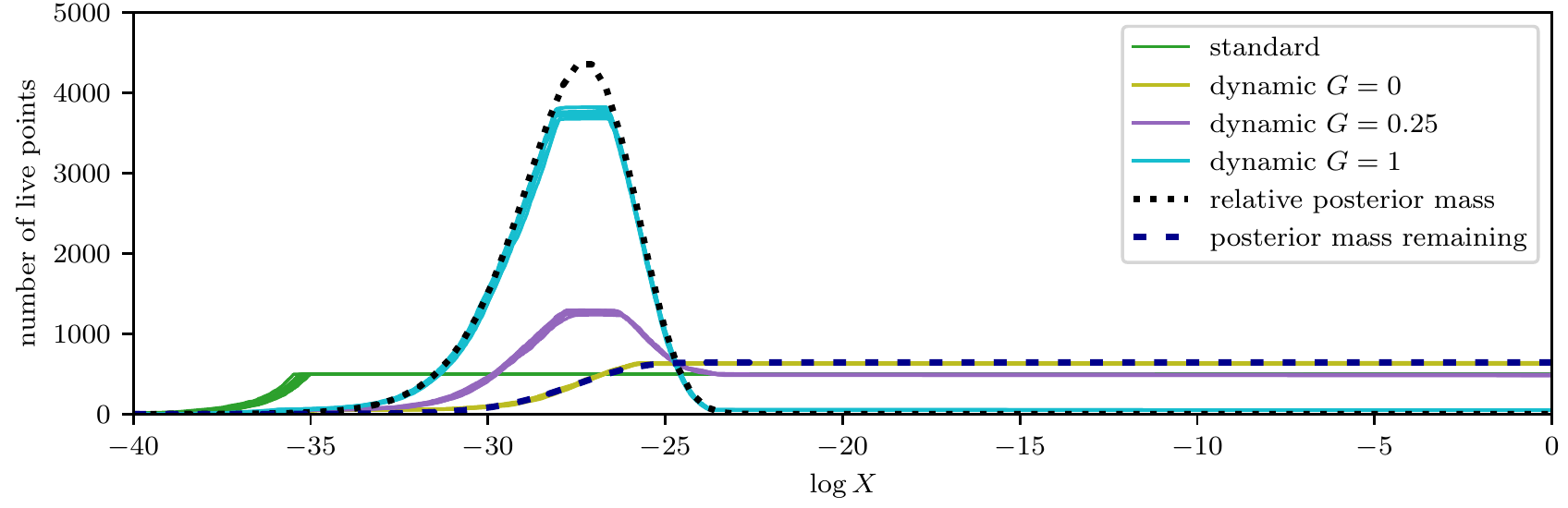}
    \caption{As in \Cref{fig:nlive_gaussian} but with a 10-dimensional exponential power likelihood~\eqref{equ:exp_power} with $b=2$.
}\label{fig:nlive_exp_power_2}
\end{figure*}

\begin{figure*}
	\centering
    \includegraphics{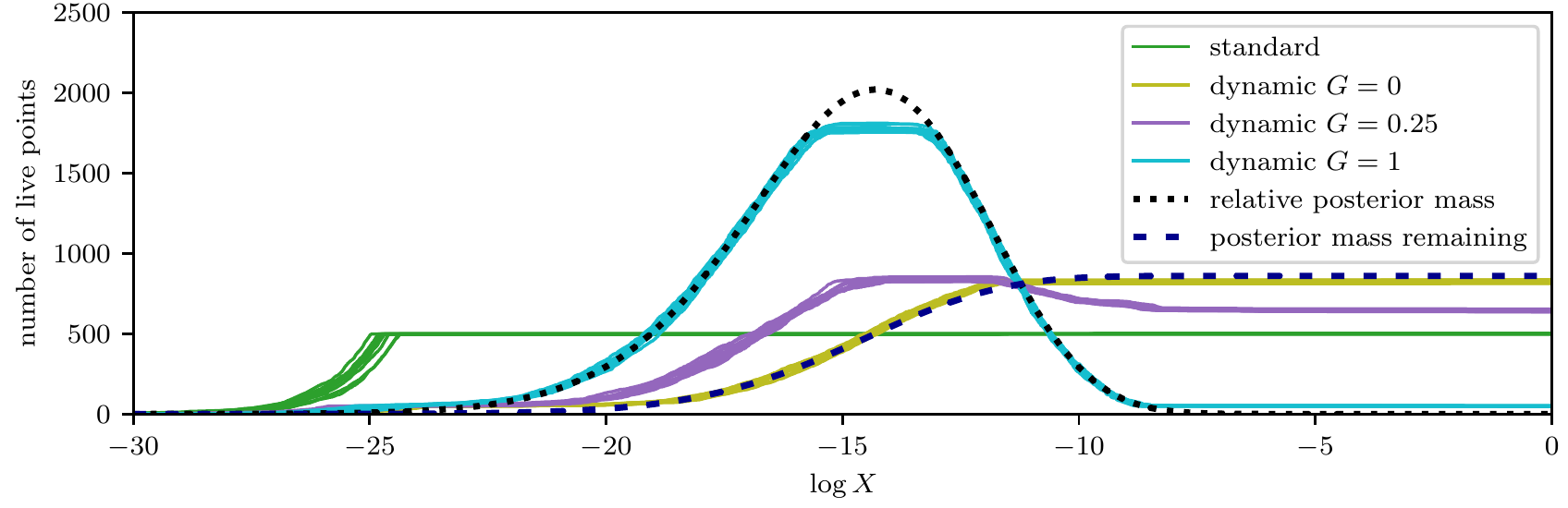}
    \caption{As in \Cref{fig:nlive_gaussian} but with a 10-dimensional exponential power likelihood~\eqref{equ:exp_power} with $b=\frac{3}{4}$.
}\label{fig:nlive_exp_power_0_75}
\end{figure*}

\Cref{fig:nlive_exp_power_2,fig:nlive_exp_power_0_75} show that the dynamic nested sampling algorithm can accurately and consistently allocate live points for these likelihoods.
\Cref{tab:dynamic_test_exp_power_2,tab:dynamic_test_exp_power_0_75} show the reduction in errors from dynamic nested sampling compared to standard nested sampling in these two cases, as measured by repeated calculations.
This corresponds to increases in efficiency~\eqref{equ:efficiency_gain} for evidence calculation ($G=0$) and parameter estimation ($G=1$) by factors of $1.25 \pm 0.04$ and up to $6.8 \pm 0.2$ respectively in the $b=2$ case, and by factors of $1.62 \pm 0.05$ and up to $3.11 \pm 0.09$ in the $b=\frac{3}{4}$ case.

\begin{table*}
\centering
    \caption{As in \Cref{tab:dynamic_test_gaussian} but with a 10-dimensional exponential power likelihood~\eqref{equ:exp_power} with $b=2$\label{tab:dynamic_test_exp_power_2}}
\begin{tabular}{llllllll}
\toprule
{} &      samples & $\log \Z$ & $\po$ & $\mathrm{median}(\thcomp{1})$ & $\mathrm{C.I.}_{84\%}(\thcomp{1})$ & $\overline{|\btheta|}$ & $\mathrm{median}(|\btheta|)$ \\
\midrule
St.Dev.\ standard             &  18,093 &                   0.228(2) &                    0.00870(9) &                           0.0110(1) &                                0.0133(1) &            0.00809(8) &                   0.0102(1) \\
St.Dev.\ $G=0$                &  18,052 &                   0.204(2) &                     0.0107(1) &                           0.0147(1) &                                0.0169(2) &            0.00917(9) &                   0.0108(1) \\
St.Dev.\ $G=0.25$             &  18,056 &                   0.228(2) &                    0.00587(6) &                          0.00777(8) &                               0.00906(9) &            0.00547(5) &                  0.00654(7) \\
St.Dev.\ $G=1$                &  18,058 &                   0.686(7) &                    0.00363(4) &                          0.00471(5) &                               0.00549(5) &            0.00338(3) &                  0.00391(4) \\
Efficiency gain $G=0$    &          &                    1.25(4) &                       0.66(2) &                             0.56(2) &                                  0.62(2) &               0.78(2) &                     0.88(2) \\
Efficiency gain $G=0.25$ &          &                    1.00(3) &                       2.20(6) &                             2.01(6) &                                  2.14(6) &               2.19(6) &                     2.41(7) \\
Efficiency gain $G=1$    &          &                   0.110(3) &                        5.7(2) &                              5.5(2) &                                   5.8(2) &                5.7(2) &                      6.8(2) \\
\bottomrule
\end{tabular}
\end{table*}

\begin{table*}
\centering
   \caption{As in \Cref{tab:dynamic_test_gaussian} but with a 10-dimensional exponential power likelihood~\eqref{equ:exp_power} with $b=\frac{3}{4}$\label{tab:dynamic_test_exp_power_0_75}}
\begin{tabular}{llllllll}
\toprule
{} &      samples & $\log \mathcal{Z}$ & $\po$ & $\mathrm{median}(\thcomp{1})$ & $\mathrm{C.I.}_{84\%}(\thcomp{1})$ & $\overline{|\btheta|}$ & $\mathrm{median}(|\btheta|)$ \\
\midrule
St.Dev.\ standard             &  12,855 &                   0.157(2) &                     0.0261(3) &                           0.0320(3) &                                0.0439(4) &             0.0545(5) &                   0.0657(7) \\
St.Dev.\ $G=0$                &  12,824 &                   0.123(1) &                     0.0283(3) &                           0.0391(4) &                                0.0487(5) &             0.0574(6) &                   0.0651(7) \\
St.Dev.\ $G=0.25$             &  12,827 &                   0.138(1) &                     0.0222(2) &                           0.0289(3) &                                0.0374(4) &             0.0454(5) &                   0.0522(5) \\
St.Dev.\ $G=1$                &  12,833 &                   0.432(4) &                     0.0160(2) &                           0.0194(2) &                                0.0266(3) &             0.0342(3) &                   0.0372(4) \\
Efficiency gain $G=0$    &          &                    1.62(5) &                       0.85(2) &                             0.67(2) &                                  0.81(2) &               0.90(3) &                     1.02(3) \\
Efficiency gain $G=0.25$ &          &                    1.30(4) &                       1.39(4) &                             1.22(3) &                                  1.38(4) &               1.44(4) &                     1.58(4) \\
Efficiency gain $G=1$    &          &                   0.132(4) &                       2.66(8) &                             2.70(8) &                                  2.71(8) &               2.54(7) &                     3.11(9) \\
\bottomrule
\end{tabular}
\end{table*}

\subsection{Gaussian mixture likelihoods}\label{app:gaussian_mix_add_tests}

\Cref{tab:gaussian_mix_rmse} shows comparisons of dynamic nested sampling results with analytically calculated values for the Gaussian mixture likelihood~\eqref{equ:gaussian_mix} with a Gaussian prior~\eqref{equ:gaussian_prior}.
The mean results are very close to the correct values, showing that there is no significant sampling bias.
As a result the root-mean-squared-errors and standard deviations are almost identical, meaning efficiency gain estimates from~\eqref{equ:efficiency_gain} can be used reliably (as for perfect nested sampling).

\begin{table*}
\centering
    \caption{Comparison of results from the nested sampling runs used in \Cref{tab:dynamic_test_gaussian_mix} with analytically calculated values for different quantities (shown in the first row).
The next 12 rows show mean, the standard deviation and root mean squared errors for the standard nested sampling runs and the dynamic nested sampling runs with $G=0$, $G=0.25$ and $G=1$.
The final 6 rows show efficiency gains calculated with the standard deviation as in~\eqref{equ:efficiency_gain}, and using the root-mean-squared-error instead of the standard deviation.
Columns show calculations of the log evidence and the mean of the first 6 parameters.
The mean dynamic nested sampling results agree closely with the analytic values, indicating that there is no significant sampling bias.
Numbers in brackets show the $1\sigma$ numerical uncertainty on the final digit.}\label{tab:gaussian_mix_rmse}
\begin{tabular}{llllllll}
\toprule
{} & $\log \mathcal{Z}$ & $\po$ & $\thmean{2}$ & $\thmean{3}$ & $\thmean{4}$ & $\thmean{5}$ & $\thmean{6}$ \\
\midrule
Analytic values                    &                   -32.3442 &                        0.3980 &                        0.3980 &                             0 &                             0 &                             0 &                             0 \\
Mean standard                  &                 -32.351(8) &                      0.397(3) &                      0.388(6) &                     0.0012(7) &                    -0.0004(7) &                     0.0005(7) &                    -0.0001(7) \\
Mean  $G=0$                    &                 -32.352(7) &                      0.393(3) &                      0.380(8) &                     0.0011(8) &                    -0.0002(8) &                     0.0005(8) &                     0.0004(9) \\
Mean  $G=0.25$                 &                 -32.336(8) &                      0.397(2) &                      0.386(5) &                    -0.0001(6) &                    -0.0007(5) &                    -0.0002(6) &                    -0.0000(6) \\
Mean  $G=1$                    &                  -32.34(2) &                      0.399(1) &                      0.385(3) &                     0.0003(4) &                    -0.0004(4) &                     0.0004(4) &                     0.0002(4) \\
St.Dev.\ standard                   &                   0.181(6) &                      0.057(2) &                      0.126(4) &                     0.0146(5) &                     0.0161(5) &                     0.0159(5) &                     0.0158(5) \\
St.Dev.\  $G=0$                     &                   0.160(5) &                      0.076(2) &                      0.176(6) &                     0.0182(6) &                     0.0178(6) &                     0.0184(6) &                     0.0192(6) \\
St.Dev.\  $G=0.25$                  &                   0.170(5) &                      0.046(1) &                      0.105(3) &                     0.0134(4) &                     0.0123(4) &                     0.0125(4) &                     0.0130(4) \\
St.Dev.\  $G=1$                     &                    0.36(1) &                      0.032(1) &                      0.069(2) &                     0.0087(3) &                     0.0089(3) &                     0.0091(3) &                     0.0086(3) \\
RMSE standard                  &                   0.181(6) &                      0.057(2) &                      0.127(4) &                     0.0147(4) &                     0.0161(5) &                     0.0159(5) &                     0.0158(5) \\
RMSE  $G=0$                    &                   0.160(5) &                      0.076(3) &                      0.177(6) &                     0.0182(6) &                     0.0178(6) &                     0.0184(5) &                     0.0192(6) \\
RMSE  $G=0.25$                 &                   0.170(5) &                      0.046(1) &                      0.106(3) &                     0.0134(5) &                     0.0123(4) &                     0.0125(4) &                     0.0130(4) \\
RMSE  $G=1$                    &                    0.36(1) &                      0.032(1) &                      0.070(2) &                     0.0086(3) &                     0.0089(3) &                     0.0091(3) &                     0.0086(3) \\
St.Dev.\ efficiency gain  $G=0$     &                     1.3(1) &                       0.56(5) &                       0.51(5) &                       0.64(6) &                       0.82(7) &                       0.75(7) &                       0.68(6) \\
St.Dev.\ efficiency gain  $G=0.25$  &                     1.1(1) &                        1.5(1) &                        1.5(1) &                        1.2(1) &                        1.7(2) &                        1.6(1) &                        1.5(1) \\
St.Dev.\ efficiency gain  $G=1$     &                    0.25(2) &                        3.3(3) &                        3.4(3) &                        2.9(3) &                        3.3(3) &                        3.1(3) &                        3.4(3) \\
RMSE efficiency gain  $G=0$    &                     1.3(1) &                       0.56(6) &                       0.51(5) &                       0.65(6) &                       0.82(7) &                       0.75(7) &                       0.68(6) \\
RMSE efficiency gain  $G=0.25$ &                     1.1(1) &                        1.5(1) &                        1.4(1) &                        1.2(1) &                        1.7(2) &                        1.6(1) &                        1.5(1) \\
RMSE efficiency gain  $G=1$    &                    0.25(2) &                        3.3(3) &                        3.3(3) &                        2.9(3) &                        3.3(3) &                        3.1(3) &                        3.4(3) \\
\bottomrule
\end{tabular}
\end{table*}

\section{Dynamic nested sampling without repeatedly restarting runs}\label{app:dyPolyChord}

This section describes the alternative dynamic nested sampling algorithm used by \dyPolyChord{} to avoid frequent resuming of the nested sampling process part way through the run.
After the initial exploratory run with $\ninit$ live points, an allocation of live points which varies with likelihood $n(\mathcal{L})$ is calculated and used to generate all the remaining samples in a single run.
The number of live points is increased during the run by sampling more than one live point from within a given iso-likelihood contour, and reduced by not replacing dead points when they are removed.
The user must specify the approximate total number of samples to be taken, $N_\mathrm{total}$, either as a constant or a function of the number of samples taken by the initial run $N_\mathrm{init}$.

The target number of live points $n(\mathcal{L})$ is calculated using importances~\eqref{equ:importance} of the dead points in the initial run; as the number of live points $\ninit$ is constant, the samples are evenly distributed in $\log X$ and the point importances are proportional to the importances of each $\log X$ region.
$n(\mathcal{L})$ is calculated piecewise at each point $i$ as
\begin{equation}
	n(\mathcal{L}_i) =
    \begin{cases}
        K \, \importance(G, i) - \ninit & \text{if $K \, \importance(G, i) > \ninit$}, \\
        0 & \text{otherwise},
    \end{cases}
    \label{equ:dpc_nlive}
\end{equation}
where $\importance(i, G)$ is point $i$'s relative importance, and each $n(\mathcal{L}_i)$ rounded to the nearest integer.
The constant $K$ is chosen so that approximately the right number of samples is taken --- i.e.\ so that $n(\mathcal{L})$ satisfies
\begin{equation}
    \int n(\mathcal{L}) \frac{\d{\log X(\mathcal{L})}}{\d{\mathcal{L}}} \d{\mathcal{L}} \approx N_\mathrm{total} - N_\mathrm{init}.
\end{equation}
If $N_\mathrm{total} \gg N_\mathrm{init}$,~\eqref{equ:dpc_nlive} allocates live points approximately in proportion to the importances calculated from the initial run.
Otherwise $n(\mathcal{L})$ is only non-zero in the region of high importance (where $\importance > \ninit / K$), and will result in approximately equal sample weights in this region in the final combined run with lower weights elsewhere.
Given the samples already taken by the initial exploratory run and the number of remaining samples available $N_\mathrm{total} - N_\mathrm{init}$,~\eqref{equ:dpc_nlive} approximately maximises the information content (Shannon entropy of the samples)~\eqref{equ:entropy}.
In practice estimates of $n(\mathcal{L})$ from~\eqref{equ:dpc_nlive} contain random noise from the stochasticity of the nested sampling algorithm.
For better results the piecewise importance function can be smoothed before calculating $n(\mathcal{L})$; by default \dyPolyChord{} uses a Savitzky-Golay filter \citep{Savitzky1964} with polynomial order 3 and window size $2 \ninit + 1$.

This procedure is set out more formally in Algorithm~\ref{alg:dypolychord}; for an example implementation see the \dyPolyChord{} package and its documentation.

\begin{algorithm}\SetAlgoLined{}
\SetKwData{Left}{left}\SetKwData{This}{this}\SetKwData{Up}{up}
\SetKwFunction{Union}{Union}\SetKwFunction{FindCompress}{FindCompress}
\SetKwInOut{Input}{Input}\SetKwInOut{Output}{Output}\SetKwInOut{op}{Other parameters}
    \Output{Samples and live points information $\mathbf{n}$.}
    \Input{Goal $G$, $\ninit$, approximate number of samples to take $N_\mathrm{total}$.}
    \BlankLine{}
    \PrintSemicolon{}
    Generate an initial nested sampling run with a constant number of live points $\ninit$\;
    calculate $n(\mathcal{L})$ from~\eqref{equ:dpc_nlive} using point importances $\importance(G, i)$ and the number of samples in the initial run $N_\mathrm{init}$\;
    perform nested sampling run with $n(\mathcal{L})$ live points, beginning by resuming initial the run at the first point where $n(\mathcal{L}_i)>0$ and terminating\footnotemark{} after the last point where $n(\mathcal{L}_i) > 0$\;
    merge the nested sampling runs generated and return the combined run.
    \vspace{0.3cm}
    \caption{The alternative dynamic nested sampling algorithm used by \dyPolyChord{}.}\label{alg:dypolychord}
\end{algorithm}%
\footnotetext{In principle $n(\mathcal{L})$ may drop to zero then, at some larger likelihood, become non-zero again --- although this is very unlikely in practice. In this case the run can terminate when $n(\mathcal{L})=0$, then be restarted at the higher likelihood when $n(\mathcal{L})$ is again non-zero by resuming the initial exploratory run at this later point.}

\section{Signal reconstruction priors and data}\label{app:fit}

\Cref{tab:fit_priors} shows the priors on the parameters used in the basis function fitting example in \Cref{sec:fit}.
The parameters of the 4 generalised Gaussians used for the true signal (from which the data was sampled) are shown in \Cref{tab:fit_data_args}.

\begin{table}
    \centering
    \caption{Priors on basis function parameters used in \Cref{sec:fit}.
Sorted priors have ordering enforced; see \citet[][Appendix A2]{Handley2015b} for more details.}%
    \label{tab:fit_priors}
    \begin{tabular}{lll}
    \toprule
    Parameter         & Prior Type          & Prior Parameters                \\
    \midrule                      
    $a$               & Sorted Exponential  & $\lambda=1                       $ \\
    $\mu$             & Uniform             & $\in [0, 1]                      $ \\
    $\sigma$          & Uniform             & $\in [0.03, 1.0]                 $ \\
    $\beta$           & Exponential         & $\lambda=0.5                     $ \\
    \bottomrule
    \end{tabular}
\end{table}

\begin{table}
    \centering
    \caption{Parameters of the sum of 4 generalised Gaussian basis functions~\eqref{equ:gg_1d} used as the true signal in \Cref{fig:fit_fgivenx}.}%
    \label{tab:fit_data_args}
    \begin{tabular}{lllll}
    \toprule
    Component    & $a$  & $\mu$ & $\sigma$  & $\beta$ \\
    \midrule
    1            & 0.2  & 0.3   & 0.5       & 5       \\
    2            & 0.4  & 0.65  & 0.07      & 2       \\
    3            & 0.6  & 0.25  & 0.1       & 6       \\
    4            & 0.9  & 0.95  & 0.1       & 6       \\
    \bottomrule
    \end{tabular}
\end{table}

\end{appendices}
\end{document}